\documentclass[aps,prd,onecolumn,preprintnumbers,nofootinbib,superscriptaddress]{revtex4-2}

\usepackage{amsmath,amssymb,amsfonts,graphicx,slashed,bbm}
\usepackage{hyperref}
\usepackage{physics}
\usepackage{color}
\usepackage{multirow}
\usepackage{caption}
\usepackage{subcaption}
\usepackage{booktabs}
\usepackage{siunitx}
\usepackage{xspace}
\usepackage{amssymb,amsmath}
\usepackage{color}
\usepackage{colortbl}
\usepackage{enumitem}
\usepackage{array}
\usepackage{braket}
\usepackage{wrapfig}
\usepackage{cancel}
\usepackage{float}
\usepackage{slashed}
\usepackage{subcaption}
\usepackage{soul}
\usepackage{cases}

\usepackage{xcolor}
\usepackage[normalem]{ulem}
\usepackage{tikz}
\usepackage{tikz-feynman}
\usepackage{natbib}
\tikzfeynmanset{compat=1.1.0}
\hypersetup{
    colorlinks=true,
    linkcolor=blue,
    citecolor=magenta,
    urlcolor=blue
}

\newcommand{\GF}{G_F}

\begin{document}

\title{Weak Annihilation Contribution to Angular Observables in $B_{c}^+\to D^{\ast+}\ell^{+}\ell^{-}$ Decays}

\author{Zohaib Aarfi}
\email{zohaib.phdphy22sns@student.nust.edu.pk}
\affiliation{Department of Physics and Astronomy ,School of Natural Sciences, National University of
Sciences and Technology, H-12, Islamabad 44000,Pakistan}
\author{Qazi Maaz Us Salam}
\email{qazimaaz92@gmail.com}
\affiliation{Department of Physics, Lahore University of Management
Sciences (LUMS), Opposite Sector U, D.H.A, Lahore 54792, Pakistan}
\affiliation{National Centre for Physics, Shahdra Valley Road, Islamabad 45320,Pakistan}
\author{Ishtiaq Ahmed}
\email{ishtiaq.ahmed@ncp.edu.pk}
\affiliation{National Centre for Physics, Shahdra Valley Road, Islamabad 45320,Pakistan}
\author{Faisal Munir Bhutta}
\email{faisal.munir@sns.nust.edu.pk}
\affiliation{Department of Physics and Astronomy ,School of Natural Sciences, National University of
Sciences and Technology, H-12, Islamabad 44000,Pakistan}
\author{M.Ali Paracha}
\email{aliparacha@sns.nust.edu.pk (corresponding author)}
\affiliation{Department of Physics and Astronomy ,School of Natural Sciences, National University of
Sciences and Technology, H-12, Islamabad 44000,Pakistan}

\begin{abstract}
We analyze the rare semileptonic decays $B_{c}^+ \to D^{\ast+}(\to P_1 P_2)\ell^{+}\ell^{-}$, with $P_1 P_2 = D^+ \pi^0$ or $D^0 \pi^+$, and $\ell=\mu, \tau$. We focus on the impact of weak annihilation contributions alongside penguin, box, and long-distance effects. Using the effective Hamiltonian for $b \to d \ell^+ \ell^-$ transitions and $B_c \to D^{*}$ form factors from covariant confined quark model inputs, we compute the differential branching ratios, forward-backward asymmetry, longitudinal helicity fraction of the $D^{\ast}$, and various normalized angular coefficients. The results of the observables show that weak annihilation effects are sizable, particularly at low $q^2$, significantly modifying several observables and shifting zero-crossings. Resonance effects dominate at high $q^2$, restricting reliable analysis windows. We conclude that the inclusion of weak annihilation is essential for precise Standard Model predictions and for isolating possible New Physics effects in $B_c^+ \to D^{*+} \ell^+ \ell^-$ decays.

\end{abstract}

\maketitle
\tableofcontents
\newpage

\section{Introduction}
The study of rare semileptonic $B$-meson decays plays a central role in testing the Standard Model (SM) and probing New Physics (NP) effects. These decays are flavor-changing neutral current (FCNC) processes based on Glashow-Iliopoulos-Maiani (GIM) \cite{Glashow:1970gm} mechanism in the SM and are forbidden at tree level but can be induced at loop level by electro-weak diagrams. Since the SM contributions are suppressed, the NP effects can give relatively large effects. In the $B_c$ decays, strong and electromagnetic decay channels are forbidden due to explicit flavors and the mass relationship $M(B_c) < M(B) + M(D)$, while the weak decay channels are allowed \cite{Ju:2013oba}. This provides us a cleaner laboratory to study weak interactions. In the $B_c$ meson both the b- and c-quark can decay, or they may both participate in a single weak process. This leads to a richer spectrum of decay channels and generally a larger available phase-space than for B or D mesons. The mass spectra of $c \bar{c}$ and $b \bar{b}$ have been studied both experimentally and theoretically due to their profound effects on $B$ meson decays. 

With the unprecedented statistics collected at the LHCb experiment, studies of rare $B_c$ decays have become feasible. Using the full Run~1 and Run~2 dataset corresponding to $9\,\mathrm{fb}^{-1}$, LHCb has recently searched for the rare decay $B_c^+ \to \pi^+ \mu^+ \mu^-$~\cite{LHCb:2023waf} and previously observed the hadronic mode $B_c^+ \to J/\psi\,\pi^+\pi^-\pi^+$~\cite{LHCb:2012ag}, paving the way for future exploration of the $B_c \to D_{s,d}^{(*)}\ell^+\ell^-$ (we will omit + sign from the $B_c$ and $D^*$ superscripts from hereon for brevity) channels. The rare decays $b \to s(d) \ell \bar{\ell}$ and $b \to s(d) \gamma$ have been focused, over the last two decades among the $B_c$ meson transitions. The study of the rare decays will be helpful to test the precision in SM and NP effects indirectly. The semi-leptonic decays $b \to s(d)\ell \bar{\ell}$ are much favored to study NP because of the large number of observables associated with them, namely the differential branching fraction, the forward-backward asymmetry, the longitudinal polarization of the final vector meson, the leptonic longitudinal polarization asymmetry and other angular observables. Further, LHCb collaboration has also provided the relative ratios for $b \to d/b \to s$ transition in the channels
$B(B^+ \to \pi^+\mu^+\mu^-)/B(B^+ \to K^+\mu^+\mu^-)$ and $B(B^0 \to K^{0*}\mu^+\mu^-)/B(B^0 \to K^{0*}\mu^+\mu^-)$ \cite{LHCb:2012de, LHCb:2018rym}.
In \cite{Artuso:2022new, London:2021lfn} several anomalies have been reported regarding these studies.

The LHCb collaboration has set an upper limit on 
\( f_c/f_u \times B(B^+ \to D_s^+ \mu^+ \mu^-) \) 
at the 95\% confidence level, where \( f_c \) and \( f_u \) denote the fragmentation fractions of 
\( B \)-mesons containing a \( c \) and \( u \) quark, respectively, in proton--proton collisions~\cite{LHCb:2023lyb}. 
On the theoretical side, semileptonic and rare semileptonic decays have been extensively studied using various frameworks. 
Geng \textit{et al.} investigated these decays within the light-front and constituent quark models~\cite{Geng:2001vy}, 
while Azizi \textit{et al.} employed the three-point QCD sum rules~\cite{Azizi:2008vv, Azizi:2008vy}. 
Faessler \textit{et al.} analyzed the exclusive rare decays \( B_c \to D^{(*)}\ell\ell \) in the relativistic quark model~\cite{Faessler:2002ut}, and Choi studied the corresponding transition form factors and observables using the light-front quark model~\cite{Choi:2010ha}. 
These decays have also been explored in the pQCD framework~\cite{Wang:2014yia} 
and within the single universal extra dimension scenario~\cite{Yilmaz:2012ah}. 
Moreover, rare semileptonic decays of \( B \) and \( B_c \) mesons have been analyzed using the relativistic quark model (RQM)~\cite{Ebert:2010dv}, 
whose form factors have further been utilized to study such decays in the presence of non-universal \( Z^0 \) effects~\cite{Maji:2020zlq}, 
within the two-Higgs-doublet model~\cite{Maji:2020wer}, 
and to probe possible NP effects through various observables~\cite{Dutta:2019wxo, Mohapatra:2021ynn, Zaki:2023mcw,Salam:2024nfv,Aarfi:2025qcp}.

The \( b \to s \ell \bar{\ell} \) processes have been explored widely through the \( B \to K^{(*)}\ell \bar{\ell}\) decays~\cite{CDF:1999uew, BaBar:2000jlq, Belle:2001oey, BaBar:2003szi, BaBar:2008jdv, Belle:2003ivt, Belle:2016fev, BELLE:2019xld, Belle:2021ecr, Belle:2009zue, Belle:2019oag, BaBar:2012mrf, CDF:2011buy, CMS:2015bcy, LHCb:2014vgu, LHCb:2016ykl, LHCb:2017avl, LHCb:2013ghj}. In the \( B \to K^{(*)}\ell \bar{\ell} \) processes, the \( b \to s \ell \bar{\ell} \) contribution is dominant, while the annihilation diagrams are CKM suppressed \( |V_{ub}^{*}V_{us}|/|V_{ts}^{*}V_{tb}| \sim \lambda^2 \) and the spectator scattering effect is at the next \( \alpha_s \) order. As shown in \cite{Ebert:2010dv, LHCb:2012juf, LHCb:2012ag}, by including only the \( b \to s \ell \bar{\ell} \) contribution in the SM, the differential branching fractions, the forward-backward asymmetries and the longitudinal polarizations of the final vector mesons are in agreement with the experimental data.

However, in \( B_c \to D_{s,d}^{(*)} \ell \bar{\ell} \) rare decays, the situation is different. The annihilation diagrams in \( B_c \to D_{s,d}^{(*)}\ell\bar{\ell} \) rare decays are CKM allowed \( |V_{cb}^{*}V_{cs(d)}|/|V_{ts(d)}^{*}V_{tb}| \sim 1 \) and enhanced by a 3 times larger color factor. Thus, the annihilation contributions should be considered before doing any NP analysis. The effects of weak annihilation (WA) contribution are studied in \cite{Ju:2013oba} for the differential branching ratio and the forward-backward asymmetry in detail. The results are compelling for the current study, where we want to explore the effects of WA on helicity fraction and other angular observables $\langle I_i \rangle$'s. 

The WA contribution is a pre-requisite to study NP effects. As pointed out in \cite{Ahmed:2011sa, Ahmed:2011fy} with the help of the interference between the \( b \to s(d)\ell\bar{\ell} \) contribution and the SM annihilation contribution, the predictions of SM with a fourth generation (SM4) and the Super-Symmetric (SUSY) Models will deviate stronger from the SM predictions in the processes \( B_c \to D_{s}^{*} \ell\bar{\ell} \) than in the processes \( B \to K^{*} \ell\bar{\ell} \). Furthermore, in \( B_c \) rare decays, the particles beyond SM may contribute not only to the \( b \to d \ell\bar{\ell} \) transitions but also to the annihilation diagrams. NP can affect the WA channel of by introducing new four-quark, scalar, vector, or tensor interactions; such as from Z$^\prime$, leptoquark, or right-handed current models that modify the effective Wilson Coefficients (WCs), alter the helicity and interference structures, and thereby impact branching ratios, polarization observables, and lepton-flavor universality ratios. If so, this double-contribution mechanism may make the NP effect more evident \cite{Ju:2013oba}.

The short- and the long-distance (LD) contributions have been studied in \cite{Ebert:2010dv, Faessler:2002ut, Geng:2001vy, Choi:2010ha, Wang:2010je, Azizi:2008vy, Azizi:2008vv}. More updated studies for the short- and long-distance contributions are presented in \cite{Isidori:2024lng, LHCb:2024onj, Bordone:2024hui, Mahmoudi:2024zna}. For the present study, we have ignored the non-factorizable contributions, such as the long-distance charm-loop corrections, as their impact on the angular observables is expected to be negligible, similar to ratios \cite{Capdevila:2017ert}. Further, the color-favored contributions are almost three times larger than the color-suppressed ones.  

To study the $B_{c} \to D_{s,d}^{(*)} \ell\bar{\ell}$ decays, we can employ the Operator Product Expansion (OPE) framework along with the factorization approach. In this formalism, the decay amplitude factorizes into two distinct components: short-distance WCs (though additional long-distance contributions may appear if resonance cascade processes are involved) and long-distance hadronic matrix elements, which represent operator expectation values between the initial and final states. Within the Renormalization Group Equation (RGE) framework, the WCs can be computed perturbatively with next-to-leading (NLO) or even next-to-next-to-leading (NNLO) order QCD corrections~\cite{Buchalla:1995vs, Misiak:2004ew}. However, evaluating the hadronic matrix elements presents a more challenging non-perturbative problem, requiring the use of model-dependent methods for their calculation.

The nonperturbative QCD dynamics of the rare decay processes 
\(B_c \to D_{s,d}^{*}\ell^+\ell^-\) are encoded in the hadronic matrix elements of the form  \(\langle D_{s,d}^{*}(p',\varepsilon)|\,\bar{q}\,\Gamma\,b\,|B_c(p)\rangle\), which are parametrized in terms of a set of Lorentz-invariant form factors. Since these form factors cannot be computed perturbatively, several nonperturbative methods have been employed in the literature to evaluate them. One of the most widely used approaches is the three-point QCD sum rule (QCDSR) method, which incorporates both perturbative and condensate contributions and has been recently applied to the \(B_c \to D_{s,d}^{(*)}\) transitions with updated input parameters~\cite{Wu:2024dzn}. Another well-established framework is the covariant light-front quark model (CLFQM), which provides the full \(q^2\)-dependence of the form factors and is particularly suited for describing heavy-to-heavy meson transitions~\cite{Hazra:2023zdk}. Earlier analyses of these form factors were also performed within the perturbative QCD (pQCD) approach based on the $k_T$ factorization theorem~\cite{Rui:2011qc}, as well as via traditional three-point QCD sum rules~\cite{Khosravi:2008jw}. Comparisons among these different methods show consistent qualitative behavior, though quantitative differences remain due to model assumptions and truncation uncertainties. In our study we take the form factors from a recent study by Ivanov, Pandya, Santorelli and Soni (IPSS) \cite{Ivanov:2024iat}. IPSS have derived the form factors for the transition $B_c^+ \to D^{(*)+}$ using covariant confined quark model (CCQM). 
The CCQM provides a coherent relativistic framework for computing hadronic transition form factors, offering several advantages over alternative quark models and nonperturbative methods. Owing to its manifest Lorentz covariance, the CCQM ensures a consistent treatment of the full kinematic region, which is particularly important for processes involving substantial recoil. The incorporation of infrared confinement directly into the quark propagators eliminates unphysical quark production thresholds and enables a seamless evaluation of form factors throughout the entire physical \(q^{2}\) domain without the need for analytic continuation between spacelike and timelike regions~\cite{Ivanov:2016qtw}. The model further relies on a small and phenomenologically well-constrained set of parameters determined from independent observables, leading to a high degree of internal consistency and predictive power~\cite{Ganbold:2014pua,Issadykov:2017wlb}. In contrast to approaches that treat mesons and baryons separately or require process-dependent modeling, the CCQM offers a unified quark-loop formulation applicable to both mesonic and baryonic transitions, thereby facilitating systematic comparisons across different decay channels. Its covariant loop integrals also naturally yield both invariant and helicity form factors, providing direct input for contemporary angular analyses of heavy-flavor decays and reducing model-dependent uncertainties~\cite{Gutsche:2014zna}. A very recent study by Dey and Nandi (DN) \cite{Dey:2025xjg} is also used to make comparison, in which the authors have employed the pQCD to extract the full kinematic range form factors for the penguin and box diagrams.  

This paper is organized as follows. In section II we present the theoretical framework, in which we describe the amplitudes for the different decay diagrams, matrix elements for the decay and finally introduce the physical observables. Then in section III we give a detailed numerical analysis of all the observables, where we discuss the impact of the weak annihilation amplitude on the observables in low and high bins. Finally in section IV we conclude our work.

\section{Theoretical Framework}

In this section, we present the weak EFT framework to study the $B_{c}\to D^{\ast}\ell^{+}\ell^{-}$ decay, driven by the quark level $b\to d\ell^{+}\ell^{-}$ transition. The EFT is described by an effective Hamiltonian that can be used to compute the four-fold angular distribution of the decay $B_{c} \to D^{\ast}(\to P_1 P_2)\ell^{+}\ell^{-}$, where $P_1 P_2 = D^+ \pi^0$ or $D^0 \pi^+$. From this angular distribution, we extract the $q^{2}$-dependent physical observables.

In this work, we investigate the semileptonic $B_{c} \to D^{\ast}(\to P_1 P_2)\ell^{+}\ell^{-}(P_1 P_2 = D^+ \pi^0$ and $D^0 \pi^+,~\ell=\mu,\tau)$ decay in the SM. The dominant contributions to this process arise from penguin diagrams and weak annihilation diagrams. Both diagrams will be used to compute the physical observables. An important aspect of this paper is to incorporate the LD resonance effects, as these contributions play a significant role in testing the SM and probing possible physics beyond the SM. 
\subsection{Penguin Box and weak Annihilation  Amplitude for $B_{c}\to  D^{\ast}(\to P_{1}P_{2})\ell^{+}\ell^{-}$ decay}
At quark level the decay  $B_{c}\to D^{\ast}\ell^{+}\ell^{-}$ is governed by $b\to d \ell^{+}\ell^{-}$ FCNC transition. The low energy effective Hamiltonian in the SM is written as ~\cite{Bause:2022rrs},

\begin{eqnarray}
\mathcal{H}_{\text{eff}}=-\frac{4G_{F}}{\sqrt{2}}V_{tb}V^{\ast}_{td}\left[\mathcal{H}^{(t)}_{\text{eff}}+\lambda_{u}^{(d)}\mathcal{H}^{(u)}_{\text{eff}}\right]+\text{h.c.},\label{Heff}
\end{eqnarray}

where  $G_{F}$ is the Fermi coupling constant, $V_{tb}V^{\ast}_{td}$ are the corresponding CKM factors and $\lambda_{u}^{(d)}=V_{ub}V^{\ast}_{ud}/V_{tb}V^{\ast}_{td}$ is the ratio of the CKM factors. The explicit form of $\mathcal{H}^{(t)}_{\text{eff}}$ and $\mathcal{H}^{(u)}_{\text{eff}}$ can be written as,
\begin{eqnarray}
\mathcal{H}^{(t)}_{\text{eff}}=C_{1}O^{c}_{1}+C_{2}O^{c}_{2}+\sum_{i=3}^{6}C_{i}O_{i}+
\sum_{i=7,9,10}C_{i}O_{i},\label{Heff1}
\end{eqnarray}
and
\begin{eqnarray}
\mathcal{H}^{(u)}_{\text{eff}}=C_{1}(O^{c}_{1}-O^{u}_{1})+C_{2}(O^{c}_{2}-O^{u}_{2}).\label{Heff2}
\end{eqnarray}

\begin{figure}[htbp]
    \centering
    \begin{subfigure}[b]{0.40\linewidth}
    \centering
    \includegraphics[width=\linewidth]{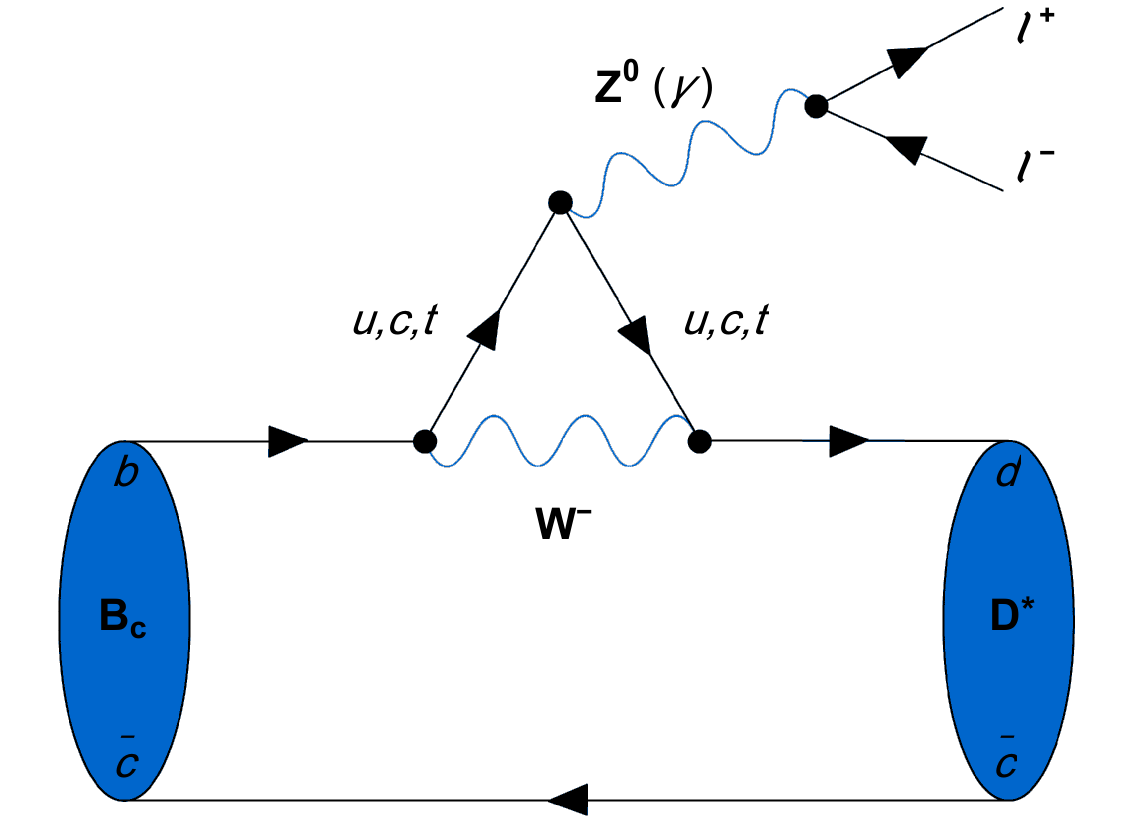}
    \caption{Penguin}
\end{subfigure}
\hspace{0.05\linewidth}
\begin{subfigure}[b]{0.40\linewidth}
    \centering
    \includegraphics[width=\linewidth]{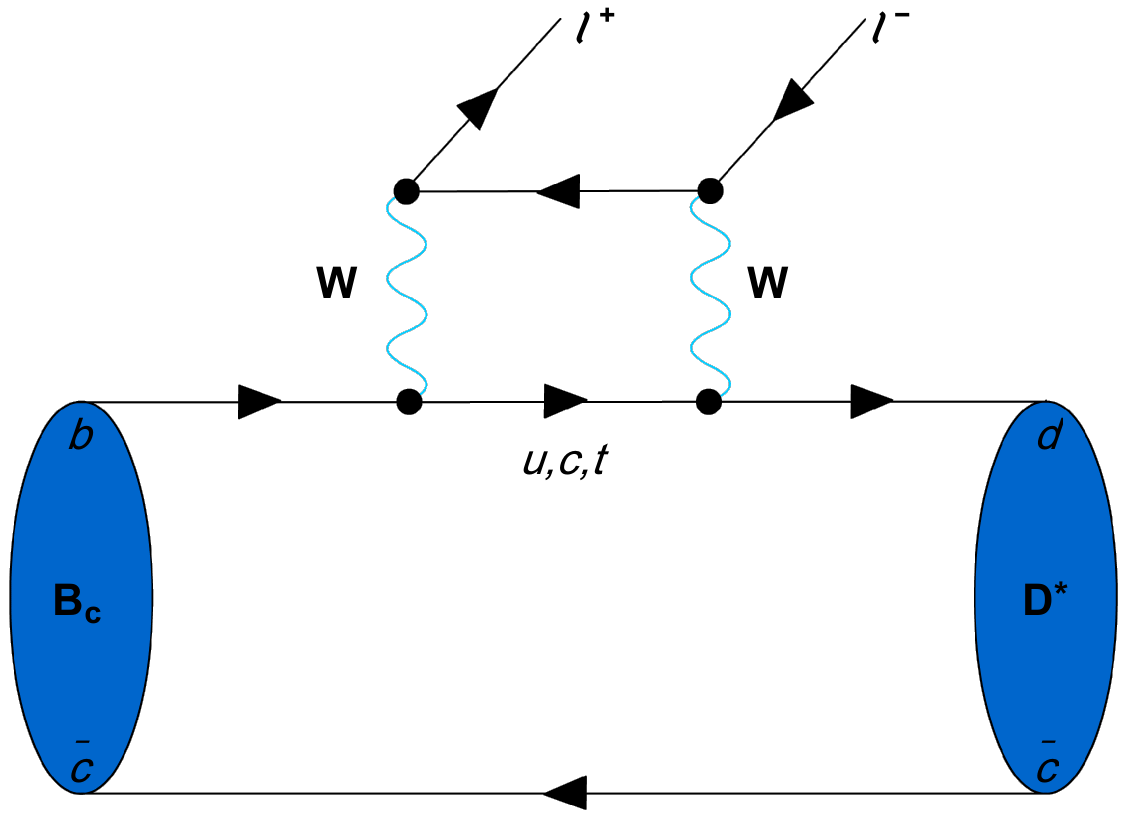}
    \caption{Box}
\end{subfigure}

    \vskip\baselineskip
    \begin{subfigure}[b]{0.40\linewidth}
        \centering
        \includegraphics[width=\linewidth]{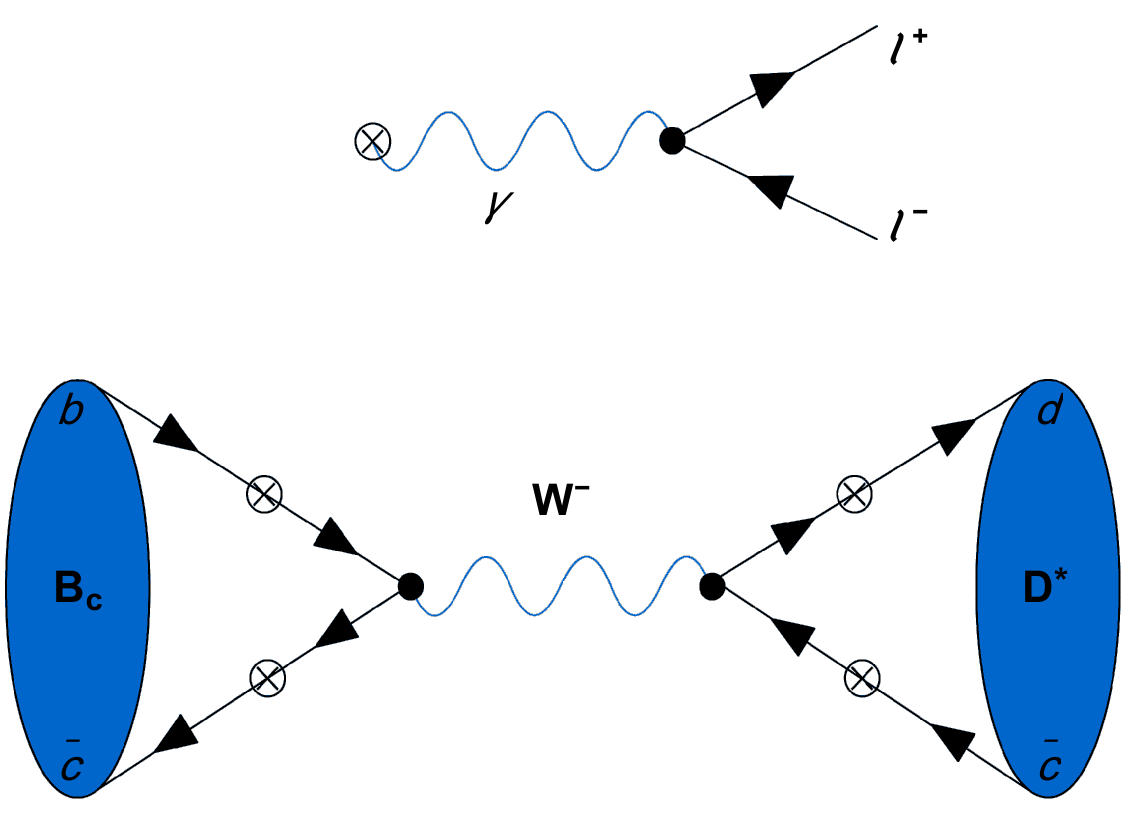}
        \caption{Weak Annihilation}
    \end{subfigure}
    \hspace{0.05\linewidth}
    \begin{subfigure}[b]{0.40\linewidth}
        \centering
        \includegraphics[width=\linewidth]{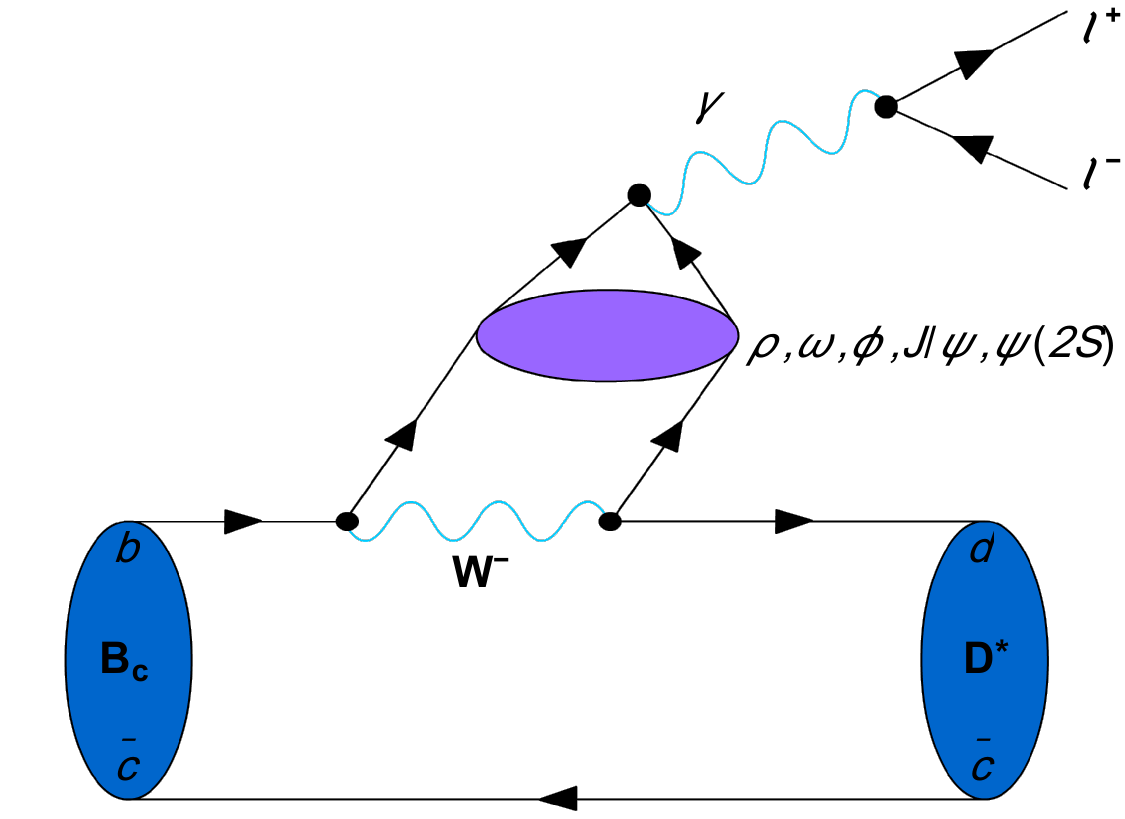}
        \caption{Color Suppressed}
    \end{subfigure}

    \caption{Feynman diagrams contributing to the process.}
    \label{fig:feynman_diagrams}
\end{figure}

In Eqs.~(\ref{Heff1}) and (\ref{Heff2}), the short distance contributions are encoded in the SM WCs $(C_{i})$, whereas the long distance effects are hidden in the local 4-fermion operators and explicitly can be written as,  
\begin{align}
O_{7} &=\frac{e}{16\pi ^{2}}m_{b}\left( \bar{d}\sigma_ {\mu \nu }P{R}b\right) F^{\mu \nu }, \\
O_{9} &=\frac{e^{2}}{16\pi ^{2}}(\bar{d}\gamma_ {\mu }P{L}b)(\bar{\ell}\gamma^{\mu }\ell), \\
O_{10} &=\frac{e^{2}}{16\pi ^{2}}(\bar{d}\gamma_ {\mu }P{L}b)(\bar{\ell} \gamma ^{\mu }\gamma ^{5}\ell).
\end{align}
$F^{\mu\nu}$ is the electromagnetic field strength tensor. It is important to mention here that $m_{b}$ appearing in the definition of the electromagnetic operator $O_{7}$ is the running $b$ quark mass and is evaluated in the $\overline{\text{MS}}$ scheme. 

Using the effective Hamiltonian given in Eq.~(\ref{Heff1}), the penguin box (PB) amplitude for the $B_{c}\to D^{\ast}\ell^{+}\ell^{-}$ decay can be expressed as,
\begin{eqnarray}
\mathcal{M}^{PB}\left(B_{c}\to D^{\ast}\ell^{+}\ell^{-}\right)=\frac{G_{F}\alpha}{2\sqrt{2}\pi}V_{tb}V^{\ast}_{td}\Big\{T^{1,PB}_{\mu}(\bar{\ell}\gamma^{\mu}\ell)
+T^{2,PB}_{\mu}(\bar{\ell}\gamma^{\mu}\gamma^{5}\ell)\Big\},\label{Amp1}
\end{eqnarray}
where
\begin{eqnarray}
T^{1,PB}_{\mu}&=&C_{9}^{\,eff}\Big\langle D^{\ast}(k,\overline{\epsilon})|\bar d\gamma_{\mu}(1-\gamma_{5})b|B_{c}(p)\Big\rangle\notag\\
&-&\frac{2m_{b}}{q^{2}}C_{7}^{\,eff}
\Big\langle D^{\ast}(k,\overline{\epsilon})|\bar d i\sigma_{\mu\nu}q^{\nu}(1+\gamma_{5})b|B_{c}(p)\Big\rangle,\label{Amp1a}
\\
T^{2,PB}_{\mu}&=&C_{10}\Big\langle D^{\ast}(k,\overline{\epsilon})|\bar d\gamma_{\mu}(1-\gamma_{5})b|B_{c}(p)\Big\rangle.\\
\label{Amp1b}
\end{eqnarray}
The explicit form of WCs $C_{7}^{eff}(q^{2})$ and $C_{9}^{eff}(q^{2})$ \cite{Bobeth:1999mk,Beneke:2001at,Asatrian:2001de,Asatryan:2001zw,Greub:2008cy,Du:2015tda}, that contain the factorizable contributions from current-current, QCD penguins and chromomagnetic dipole operators $O_{1-6,8}$ is given in Appendix \ref{appB}.  

For WA diagram, the amplitude for $B_{c}\to D^{\ast}\ell^{+}\ell^{-}$ decays can be written by using the Naive factorization as \cite{Ju:2013oba}

\[
\mathcal{M}_{Ann} 
= V_{cb} V_{cd}^{*} 
\frac{i \alpha}{q^2}
\frac{G_F}{2 \sqrt{2} \pi}
\left( \frac{C_1}{N_c} + C_2 \right)
\mathcal{T}_{\text{ann}}^{\mu} \, \bar{\ell} \gamma_{\mu}\ell,
\]
where,
\[
\mathcal{T}_{\text{ann}}^{\mu} = \mathcal{T}_1^{\mu} + \mathcal{T}_2^{\mu} + \mathcal{T}_3^{\mu} + \mathcal{T}_4^{\mu}.
\]

The hadronic matrix elements for the WA diagram can explicitly be written as,

\begin{align}
\mathcal{T}_1^{\mu} &= (-8\pi^2) 
\langle D^{*} | \bar{d} \gamma_{\alpha} (1 - \gamma_5) c | 0 \rangle
\langle 0 | \bar{c} \gamma^{\alpha} (1 - \gamma_5) 
\frac{1}{\slashed{p}_{q_1} - m_{q_1} + i\epsilon}
\left(-\frac{1}{3}\right) \gamma^{\mu} b | B_c \rangle,\\
\mathcal{T}_2^{\mu} &= (-8\pi^2) 
\langle D^{*} | \bar{d} \gamma_{\alpha} (1 - \gamma_5) c | 0 \rangle \langle 0 |
\left( \frac{2}{3} \right) \gamma^{\mu} 
\frac{1}{\slashed{p}_{q_2} - m_{q_2} + i\epsilon}
\gamma^{\alpha} (1 - \gamma_5) b | B_c \rangle,\\
\mathcal{T}_3^{\mu} &= (-8\pi^2) 
\langle D^{*} | \bar{d}
\left(-\frac{1}{3}\right) \gamma^{\mu}
\frac{1}{\slashed{p}_{q_4} - m_{q_4} + i\epsilon}
\gamma_{\alpha} (1 - \gamma_5) c | 0 \rangle
\langle 0 | \bar{c} \gamma^{\alpha} (1 - \gamma_5) b | B_c \rangle,\\
\mathcal{T}_4^{\mu} &= (-8\pi^2) 
\langle D^{*} | \bar{d} \gamma_{\alpha} (1 - \gamma_5)
\frac{1}{\slashed{p}_{q_3} - m_{q_3} + i\epsilon}
\left(\frac{2}{3}\right) \gamma^{\mu} c | 0 \rangle
\langle 0 | \bar{c} \gamma^{\alpha} (1 - \gamma_5) b | B_c \rangle.
\end{align}
Where $p_{q_{1},q_{2},...,q_{4}}$ and $m_{q_{1},q_{2},...,q_{4}}$ represent the momenta and masses of quarks respectively.

The LD effects in this work are induced by the resonance decays, i.e. $B_{c}\to D^{\ast}V\to D^{\ast}\ell^{+}\ell^{-}$ and its contribution can be described by the relation $B(B_{c}\to D^{\ast}\ell^{+}\ell^{-})\sim B(B_{c}\to D^{\ast}V)\times B(V\to\ell^{+}\ell^{-})$ approximately \cite{Ju:2013oba}, where $(V=J/\psi,\psi(2s),\rho,\omega,\phi)$ respectively. These resonances are added in the Wilson coefficient $C^{\text{eff}}_{9}$ in the SM provided in Appendix \ref{appB} and it's explicit form can be written as,
\begin{align}
Y_{\rm res}(q^2,m_b^{pole}) &= -\frac{3\pi}{\alpha_e^2}\biggl[C(m_b^{pole}) \sum_{V_i=J/\psi,\psi(2S),\ldots}\frac{m_{V_i}B(V_i\rightarrow\ell^{+}\ell^{-})\Gamma_{V_i}}{q^2-m_{V_i}^2+i m_{V_i}\Gamma_{V_i}}\notag \\
&\quad -\lambda_{u}B(0,\hat{s})(3C_1(m_b^{pole})+C_2(m_b^{pole}))\notag \\
&\quad \times\sum_{V_j=\rho,\omega,\phi}\frac{m_{V_j}B(V_j\rightarrow\ell^{+}\ell^{-})\Gamma_{V_j}}{q^2-m_{V_j}^2+i m_{V_j}\Gamma_{V_j}}\biggr].\label{res}
\end{align}

Where \(\hat{s} = q^2/m_b^2\) and \(\hat{m}_c = m_c/m_b\) with \(m_b = 4.8 \, \text{GeV}\) and \(m_c = 1.6 \, \text{GeV}\), and \(C(m_b^{pole}) = 3C_1(m_b^{pole}) + C_2(m_b^{pole}) + 3C_3(m_b^{pole}) + C_4(m_b^{pole}) + 3C_5(m_b^{pole}) + C_6(m_b^{pole})\). At the next leading order, the WCs at the QCD renormalization scale \(m_b^{pole} = m_b\) are chosen as \(C_1 = -0.175\), \(C_2 = 1.076\), \(C_3 = 1.258\%\), \(C_4 = -3.279\%\), \(C_5 = 1.112\%\), \(C_6 = -3.634\%\), \(C_7 = -0.302\), \(C_8 = -0.148\), \(C_9 = 4.232\), and \(C_{10} = -4.410\) \cite{Buchalla:1995vs}.

\subsection{Matrix Elements for the \texorpdfstring{$B_{c}\to D^{\ast}\ell^{+}\ell^{-}$}{Bs to K* mu+ mu-} decay}\label{ffP}
For the decay $B_{c}\to D_{d}^{\ast}\ell^{+}\ell^{-}$, the transition matrix elements given in  Eqs. (\ref{Amp1a}) and (\ref{Amp1b}) can be expressed in terms of form factors relevant to vector and axial vector currents as transition form factors as,

\begin{align}
\left\langle D^\ast(k,\overline\epsilon)\left\vert \bar{d}\gamma
_{\mu }b\right\vert B_c(p)\right\rangle &=\frac{2\epsilon_{\mu\nu\alpha\beta}}
{m_{B_c}+m_{D^\ast}}\overline\epsilon^{\,\ast\nu}p^{\alpha}k^{\beta}V(q^{2}),\label{2.13a}
\\
\left\langle D^\ast(k,\overline\epsilon)\left\vert \bar{d}\gamma_{\mu}\gamma_{5}b\right\vert
B_c(p)\right\rangle &=i\left(m_{B_c}-m_{D^\ast}\right)g_{\mu\nu}\overline\epsilon^{\,\ast\nu}A_{1}(q^{2})
-\frac{iP_{\mu}(\overline\epsilon^{\ast}\cdot q)A_{2}(q^{2})}{\left(m_{B_c}+m_{D^\ast}\right)}
-\frac{iq_{\mu}(\overline\epsilon^{\,\ast}\cdot q)A_{3}(q^{2})}{\left(m_{B_c}+m_{D^\ast}\right)},\label{2.13b}
\end{align}
where $P_{\mu}=p_{\mu}+k_{\mu}$, $q_{\mu}=p_{\mu}-k_{\mu}$. We have used the $\epsilon_{0123}=+1$ convention throughout the study. The tensor form factors also contribute in this process and the matrix elements associated with tensor operators can be written as,

\begin{align}
\left\langle D^\ast(k,\overline\epsilon)\left\vert \bar{d}i\sigma
_{\mu \nu }q^{\nu }b\right\vert B_{c}(p)\right\rangle
&=-2\epsilon _{\mu\nu\alpha\beta}\overline\epsilon^{\,\ast\nu}p^{\alpha}k^{\beta}T_{1}(q^{2}),\label{FF11}\\
\left\langle D^\ast(k,\overline\epsilon )\left\vert \bar{d}i\sigma
_{\mu \nu }q^{\nu}\gamma_{5}b\right\vert B_{c}(p)\right\rangle
&=i\Big[\left(m^2_{B_c}-m^2_{D^\ast}\right)g_{\mu\nu}\overline\epsilon^{\,\ast\nu}-(\overline\epsilon^{\,\ast }\cdot q)P_{\mu}\Big]T_{2}(q^{2})+i(\overline\epsilon^{\,\ast}\cdot q)\left[q_{\mu}-\frac{q^{2} P_{\mu}}{m^2_{B_{c}}-m^2_{D^\ast}}
\right]T_{3}(q^{2}).\label{F3}
\end{align}

The form factors of weak annihilation amplitude can be expressed as,
\begin{align}
T^{\mu}_{\text{ann}}(B_{c}\to D^{\ast})=(m_{B_{c}}-m_{D^{\ast}}) \left[T_{1\text{ann}}m^{2}_{B_{c}}\overline\epsilon^{\mu}+T_{2\text{ann}}(\overline\epsilon.q)p^{\mu}+T_{3\text{ann}}(\overline\epsilon.q)k^{\mu}+\frac{1}{2}iV_{\text{ann}}\epsilon_{\mu\nu\alpha\beta}\overline\epsilon^{\nu}p^{\alpha}k^{\beta}\right],
\end{align}
where $V(q^{2}),~A_{1}(q^{2}),~A_{2}(q^{2}),~A_{3}(q^{2}),~T_{1}(q^{2}),~T_{2}(q^{2}),~T_{3}(q^{2}),~T_{1\text{ann}},~T_{2\text{ann}},~T_{3\text{ann}}$ and $V_{\text{ann}}(q^{2})$ are the form factors associated with penguin diagrams as well as the weak annihilation diagrams.

\subsection{Physical Observables for the $B_{c}\to D^{\ast}(\to P_1 P_2)\ell^{+}\ell^{-}$~decay}

\begin{figure}
    \centering
    \includegraphics[width=0.8\linewidth]{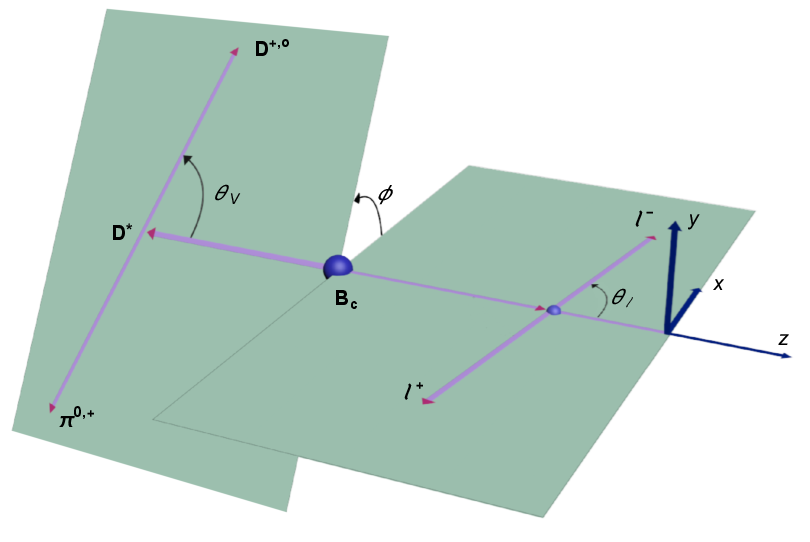}
    \caption{Kinematics of the  $B_{c}\to D^{\ast}(\to P_1 P_2)l^{+}l^{-}$ decay.}
    \label{cascadeDecay}
\end{figure}

The physical observables, we use to analyze the WA effects in $B_{c}\to D^{\ast}(\to P_1 P_2)\ell^{+}\ell^{-}$ decays are differential branching ratio $dB(B_{c}\to D^{\ast}(\to P_{1}P_{2})\ell^{+}\ell^{-})/dq^{2}$ forward backward asymmetry $\text{A}_{FB}$, longitudinal helicity fraction $f_{L}$ and normalized angular coefficients $\langle I_{n}\rangle$. These observables are computed by using angular coefficients $I_{n}$'s which appears in the formula for four fold decay distribution. The four fold differential decay rate can be expressed  in terms of the square of the momentum transfer $q^{2}$, and the kinematic angles (see Fig.~\ref{cascadeDecay}) $\theta_{\ell}$, $\theta_{V}$, and $\phi$  as

\begin{eqnarray}
 \frac{d^4\Gamma\left(B_{c}\to D^{\ast}\,(\to P_1 P_2)\ell^+\ell^-\right)}{dq^2 \ d\cos{\theta_{l}} \ d\cos {\theta}_{V} \ d\phi} &=& \frac{9}{32 \pi} B(D^{\ast}\to P_1 P_2)\bigg[I_{1s}\sin^2\theta_{V}+I_{1c}\cos^2\theta_{V}
+\Big(I_{2s}\sin^2\theta_{V}+I_{2c}\cos^2\theta_{V}\Big)\cos{2\theta_{l}}
\notag\\
&+&\Big(I_{6s}\sin^2\theta_{V}+I_{6c}\cos^2\theta_{V}\Big)\cos{\theta_{l}}+\Big(I_{3}\cos{2\phi}
+I_{9}\sin{2\phi}\Big)\sin^2\theta_{V}\sin^2\theta_{l}\notag
\\
&+&\Big(I_{4}\cos{\phi}+I_{8}\sin{\phi}\Big)\sin2\theta_{V}\sin2\theta_{l}+\Big(I_{5}\cos{\phi}+I_{7}\sin{\phi}\Big)\sin2\theta_{V}\sin\theta_{l}\bigg].
\label{fullad}
\end{eqnarray}

 The expressions of helicity amplitudes in terms of SM WCs as well as transition form factors for the process $B_{c}\to D^{\ast}$ are,

\begin{align}
\label{helicityamps}
H^{1,D^{\ast}}_t&=-i\sqrt{\frac{\lambda}{q^2}}C_{9}^{eff}A_0 -\frac{i \sqrt{\lambda} R_{PBAnn}}{4 \sqrt{q^2} M_{D^\ast}} (M_{B_c}-M_{D^\ast})\{2M^2_{D^\ast} T_{1ann} + (q^2+M^2_{B_c}-M^2_{D^\ast})T_{2ann} \\
&- (q^2-M^2_{B_c}+M^2_{D^\ast})T_{3ann} \},\notag
\\
H^{2,D^{\ast}}_t&=-i\sqrt{\frac{\lambda}{q^2}}C_{10}A_0,\notag
\\
H^{1,D^{\ast}}_{\pm}&=-i\left(M^2_{B_c}-M^2_{D^\ast}\right)\Big[C_{9}^{eff}
\frac{A_{1}}{\left(M_{B_c}-M_{D^\ast}\right)}
+\frac{2m_{b}}{q^{2}}C_{7}^{eff}T_{2}\Big]
\pm i\sqrt{\lambda}\Big[C_{9}^{eff}
\frac{V}{\left(M_{B_c}+M_{D^\ast}\right)} \notag \\
&+\frac{2m_{b}}{q^{2}}C_{7}^{eff}T_{1}\Big] - i  R_{PBAnn}\big[ M^2_{B_c}(M_{B_c}-M_{D^\ast}) T_{1ann} \pm  \frac{\sqrt{\lambda}}{2} (M_{B_c}-M_{D^\ast}) V_{ann}\big],\notag
\\
H^{2,D^\ast}_{\pm}&=-iC_{10}\left(M_{B_c}+M_{D^\ast}\right)
A_{1}\pm i\sqrt{\lambda}C_{10}
\frac{V}{\left(M_{B_c}+M_{D^\ast}\right)},\notag
\\
H^{1,D^\ast}_0&=-\frac{i}{2M_{D^\ast}\sqrt{q^2}}\Bigg[C_{9}^{eff}
\Big\{(M^2_{B_c}-M^2_{D^\ast}-q^2)\left(M_{B_c}+M_{D^\ast}\right)A_{1}\notag
\\
&-\frac{\lambda}{M_{B_c}+M_{D^\ast}}A_{2}\Big\}+2m_b C_{7}^{eff}\Big\{(M^2_{B_c}+3M^2_{D^\ast}-q^2)T_{2} -\frac{\lambda}{M^2_{B_c}-M^2_{D^\ast}}T_{3}\Big\}\Bigg]\\
&-\frac{i R_{PBAnn}}{4 \sqrt{q^2} M_{D^\ast}(M_{B_c}+M_{D^\ast})} (M^2_{D^\ast}-M^2_{B_c})\{2 M^2_{B_c} T_{1ann} (q^2-M^2_{B_c}+M^2_{D^\ast})\\
&- \lambda (T_{2ann} + T_{1ann})\} ,\notag \\
H^{2,D^\ast}_0&=-\frac{i}{2M_{D^\ast}\sqrt{q^2}}C_{10}
\Bigg[(M^2_{B_c}-M^2_{D^\ast}-q^2)\left(M_{B_c}+M_{D^\ast}\right)A_{1}-\frac{\lambda}{M_{B_c}+M_{D^\ast}}A_{2}\Bigg].\label{HA6}
\end{align}

Where $R_{PBAnn} = (V_{cb} V^*_{cd} /V_{tb} V^*_{td})(C_1/3+C_2)$. 

The angular coefficients given in Eqs.~(\ref{I1s})-(\ref{I6s}) can be used to compute the physical observables such as branching ratio $dB(B_{c}\to D^{\ast}\ell^{+}\ell^{-})/dq^{2}$, the forward-backward asymmetry $A_{\text{FB}}$, the longitudinal polarization of the $D^{\ast}$ meson $f_{L}$ and the normalized angular coefficients $I_{i}$. The angular coefficients $I_{i}$ which, in terms of helicity amplitudes are given as,

\begin{eqnarray}
&I_{1s}& = \frac{(2+\beta_l^2)}{2}N^2\left(|H_+^1|^2+|H_+^2|^2+|H_-^1|^2+|H_-^2|^2\right)\notag
\\&+&\frac{4m_l^2}{q^2}N^2\left(|H_+^1|^2-|H_+^2|^2+|H_-^1|^2-|H_-^2|^2\right),\label{I1s}\\
&I_{1c}& = 2N^2\left(|H_0^1|^2+|H_0^2|^2\right)+\frac{8m_l^2}{q^2}N^2\left(|H_0^1|^2-|H_0^2|^2+2|H_t^2|^2\right),\\
&I_{2s}& = \frac{\beta_l^2}{2}N^2\left(|H_+^1|^2+|H_+^2|^2+|H_-^1|^2+|H_-^2|^2\right),\\
&I_{2c}& = -2\beta_l^2N^2\left(|H_0^1|^2+|H_0^2|^2\right),\\
&I_{3}&=-2\beta_l^2N^2\bigg[\mathcal{R}e\left(H_+^{1}H_-^{1\ast}+H_+^{2}H_-^{2\ast}\right)\bigg],\\
&I_{4}&=\beta_l^2N^2\bigg[\mathcal{R}e\left(H_+^{1}H_0^{1\ast}+H_-^{1}H_0^{1\ast}\right)
+\mathcal{R}e\left(H_+^{2}H_0^{2\ast}+H_-^{2}H_0^{2\ast}\right)\bigg],\\
&I_{5}&=-2\beta_lN^2\bigg[\mathcal{R}e\left(H_+^{1}H_0^{2\ast}-H_-^{1}H_0^{2\ast}\right)
+\mathcal{R}e\left(H_+^{2}H_0^{1\ast}-H_-^{2}H_0^{1\ast}\right)\bigg],\\
&I_{6s}&=-4\beta_lN^2\bigg[\mathcal{R}e\left(H_+^{1}H_+^{2\ast}-H_-^{1}H_-^{2\ast}\right)\bigg],\label{I6s}\\
&I_{6c}&=0,\\
&I_{7}&=-2\beta_lN^2\bigg[\mathcal{I}m\left(H_0^{1}H_+^{2\ast}+H_0^{1}H_-^{2\ast}\right)
+\mathcal{I}m\left(H_0^{2}H_+^{1\ast}+H_0^{2}H_-^{1\ast}\right)\bigg],\\
&I_{8}&=\beta_l^2N^2\bigg[\mathcal{I}m\left(H_0^{1}H_+^{1\ast}-H_0^{1}H_-^{1\ast}\right)
+\mathcal{I}m\left(H_0^{2}H_+^{2\ast}-H_0^{2}H_-^{2\ast}\right)\bigg],\\
&I_{9}&=2\beta_l^2N^2\bigg[\mathcal{I}m\left(H_+^{1}H_-^{1\ast}+H_+^{2}H_-^{2\ast}\right)\bigg]. 
\end{eqnarray}

Here, 
\begin{eqnarray}\label{Norm}
N=V_{tb}V^{\ast}_{td}\Bigg[\frac{G_{F}^2\alpha^2}{3.2^{10} \pi^5 M_{B_c}^{3}} q^2\sqrt{\lambda}\beta_l\Bigg]^{1/2},
\end{eqnarray}
with $\lambda\equiv \lambda(M^2_{B_c}, M^2_{D^{\ast}}, q^2)$ and $\beta_l=\sqrt{1-4m_l^2/q^2}$.  

We now present the formulas used for computing the physical observables. 

\begin{itemize}
\item Differential decay rate: In terms of angular coefficients, the differential decay rate for the decay $B_{c}\to D^{\ast}(\to P_1 P_2)\ell^{+}\ell^{-}$
     can be expressed as,
    \begin{equation}
        \frac { \mathrm { d } \Gamma ( B_c \rightarrow D^* ( \rightarrow P_1 P_2 ) \ell ^ { + } \ell ^ { - } ) } { \mathrm { d } q ^ { 2 } } = B ( D^* \rightarrow P_1 P_2 ) \frac { 1 } { 4 } ( 3 I _ { 1c } + 6 I _ { 1s } - I _ {2c} - 2 I _ { 2s }). \label{Br1}
    \end{equation}

    \item {Lepton forward-backward asymmetry:} The lepton forward-backward asymmetry for the $B_{c}\to D^{\ast}(\to P_1 P_2)\ell^{+}\ell^{-}$ decay in terms of angular coefficients $I_{i}$ can be written as,
    \begin{equation}
        A_{\mathrm{FB}}(q^{2}) = \frac{6I_{6s}}{2(3I_{1c} + 6I_{1s} - I_{2c}- 2I_{2s})}.\label{AFB1}
    \end{equation}

    \item {Longitudinal helicity fraction:} The longitudinal helicity fraction for the $B_{c}\to D^{\ast}(\to P_1 P_2)\ell^{+}\ell^{-}$ decay  in terms of angular coefficients  $I_i$ can be expressed as,
    \begin{equation}
        f_L (q^2) = \frac{3 I_{1c} - I_{2c}}{3 I_{1c} + 6 I_{1s} - I_{2c} - 2 I_{2s}}.\label{fL1}
    \end{equation}

    \item {Normalized angular coefficients:} The normalized angular coefficients are defined as,
    \begin{equation}
    \langle I_{i}\rangle = 
    \frac{
    B(D^* \to P_1 P_2) \, I_{i}
    }{
    \mathrm{d}\Gamma\left( B_c \to D^* (\to P_1 P_2)\ell^+ \ell^- \right)/\mathrm{d}q^{2}
    }.\label{angobsorig}
    \end{equation}
\end{itemize}

These angular observables can be obtained from the data through the maximum likelihood fit or via method of total and partial moments (MoM) \cite{LHCb:2018jna, Gratrex:2015hna, James:2006zz, Beaujean:2015xea}. In experiment the angular information about $B \to D^* \ell \bar{\ell}$ is extracted at the level of $I_i$'s \cite{LHCb:2015svh}, and also suggested for the analysis at the amplitude level \cite{Egede:2015kha}.
\section{Numerical Analysis}
\subsection{Input Parameters}
In this section, we analyze the weak annihilation effect along with penguin box (PB) and LD effects in $B_{c}\to D^{\ast}(\to P_{1}P_{2})\ell^{+}\ell^{-}$ decays through the differential branching ratio $dB/dq^{2}$, forward backward asymmetry $A_{FB}(q^{2})$, longitudinal helicity fractions $f_{L}$ and angular coefficients $\langle I_{i}\rangle's$ in the SM framework. To analyze such effects, one needs the relevant transition form factors as necessary input parameter. The form factors, as mentioned earlier, appearing in PB diagram were evaluated in the CCQM, and their parameterization in the whole $q^{2}$ region is given as follows~\cite{Ivanov:2024iat}.
\begin{eqnarray}
F(q^2) = \frac{F(0)}{1 - a \left( \frac{q^2}{m_1^2} \right) + b \left( \frac{q^2}{m_1^2} \right)^2} ,\label{FF}
\end{eqnarray}
the value of  form factors at $q^{2}=0$ and double pole parameters $a$ and $b$ are presented in the Table \ref{FF table11}. 

\begin{table}[htbp]
\centering
\caption{Form factors and double pole parameters appeared in Eq.~\eqref{FF}~\cite{Ivanov:2024iat}.}
\label{FF table11}
\begin{tabular}{|c|c|c|c|c|c|c|c|}
\hline
$F$ & $F(0)$ & $a$ & $b$ & $F$ & $F(0)$ & $a$ & $b$ \\
\hline

$A_1$ & $0.278 \pm 0.002$ & $1.447 \pm 0.010$ & $0.171 \pm 0.026$   & $A_2$ & $0.152 \pm 0.001$ & $2.155 \pm 0.007$ & $1.089 \pm 0.017$\\
$A_3$ & $-(0.237 \pm 0.002)$ & $2.409 \pm 0.006$ & $1.459 \pm 0.015$ & $V$ & $0.232 \pm 0.002$ & $2.392 \pm 0.006$ & $1.421 \pm 0.015$ \\
$T_1$ & $0.144 \pm 0.001$ & $2.472 \pm 0.006$ & $1.554 \pm 0.014$ & $T_2$ & $0.143 \pm 0.001$ & $1.532 \pm 0.010$ & $0.294 \pm 0.025$ \\
$T_3$ & $0.143 \pm 0.009$ & $2.147 \pm 0.007$ & $1.090 \pm 0.018$ & & & & \\
\hline
\end{tabular}
\end{table}
The calculations of the WA form factors $T^{\mu}_{\text{ann}}$ are different than that of penguin box matrix elements, which we have adopted from \cite{Ju:2013oba}. These form factors are complex and can be written as,
\begin{equation}
    \begin{aligned}
        T_{1\text{ann}}&= T_{1\text{annr}} +  iT_{1\text{anni}}, \quad T_{2\text{ann}}= T_{2\text{annr}} + i T_{2\text{anni}}\\
        T_{3\text{ann}}&= T_{3\text{annr}} + i T_{3\text{anni}}, \quad V_{\text{ann}}= V_{\text{annr}} + i V_{\text{anni}}.
    \end{aligned}
\end{equation}

The other input parameters, such as the masses of mesons, Fermi coupling constant $G_{F}$, value of CKM matrix elements and fine structure constant $\alpha$ are listed in Table~\ref{input}.
\begin{table}[h!]
    \centering
    \caption{Input parameters used in this analysis.}\label{input}
    \begin{tabular}{lcc}
        \toprule
        Quantity & Symbol & Value \\
        \midrule
        $B_c$ mass & $m_{B_c}$ & \SI{6.274}{GeV} \\
        $D^*$ mass & $m_{D^*}$ & \SI{2.010}{GeV} \\
        Fermi constant & $\GF$ & \SI{1.166e-5}{GeV^{-2}} \\
        CKM element & $|V_{cb}|$ & 0.041 \\
        Fine structure constant & $\alpha$ & 1/137 \\
        \bottomrule
    \end{tabular}
\end{table}

\subsection{Analysis of Physical observables in $B_c \to D^{*}(\to D^{+} \pi^{0}(D^{0} \pi^{+}))\ell^+\ell^- $ Decays.}
We now present the phenomenological analysis of the above mentioned physical observables for  $B_c \to D^*(\to D^{+,0} \pi^{0,+})\ell^+\ell^- $ ($\ell = \mu,\tau$), decays. We analyze the observables in both low and high $q^{2}$ for $B_c \to D^*(\to D^{+,0} \pi^{0,+})\mu^+\mu^- $ decay. The predicted numerical values of the said observables in different $q^{2}$ bins in SM are presented in Table-\ref{binned}. Furthermore in Figs. \ref{Fig1}-\ref{Fig4} we plot the physical observables as a function of $q^{2}$. The first two columns in each figure correspond to low and high $q^{2}$ region for the case $\ell=\mu$ as a final state lepton. In the third column we plotted the same observables in high  bin for $\ell=\tau$. The PB contributions are represented by the green bands and the effects of LD+PB are given in red bands. The Effects of WA+PB are presented in brown bands and finally the overall combined effects PB+WA+LD are presented in blue bands. These bands contain the uncertainties of all the form factors and the double pole parameters as well.

\subsubsection{Branching Ratio}

\begin{figure}[H]
\centering
\includegraphics[width=2.2in,height=1.36in]{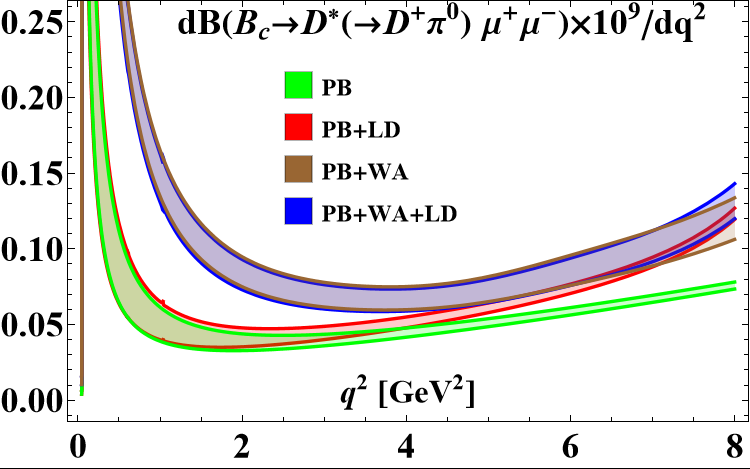}
\includegraphics[width=2.2in,height=1.36in]{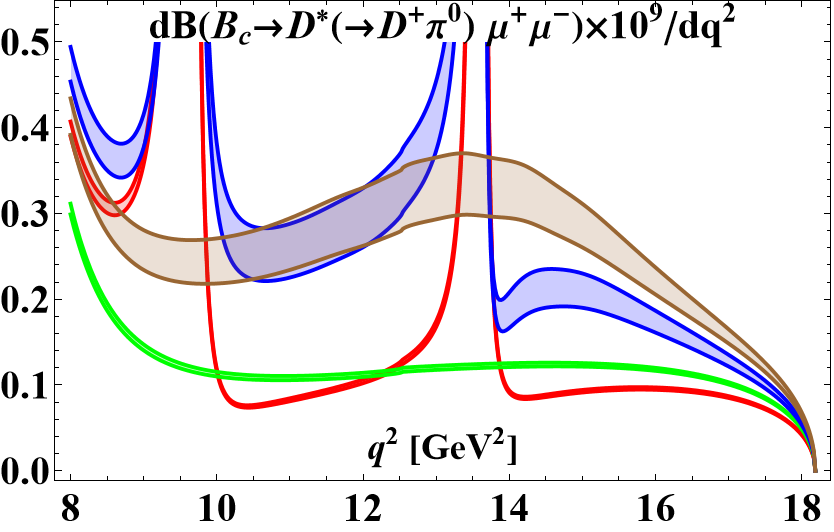}
\includegraphics[width=2.2in,height=1.36in]{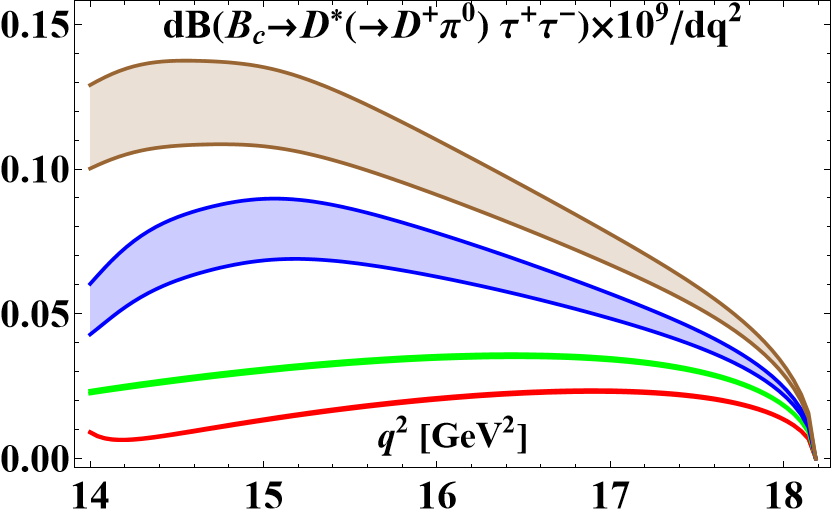}

\includegraphics[width=2.2in,height=1.36in]{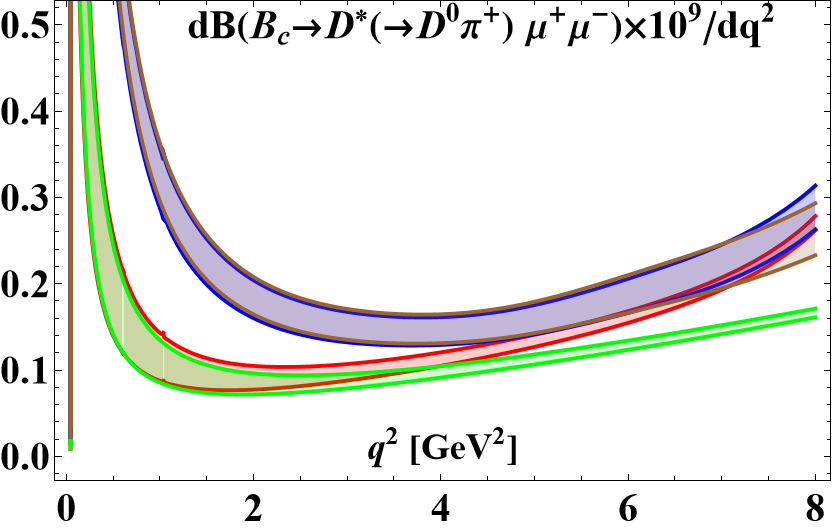}
\includegraphics[width=2.2in,height=1.36in]{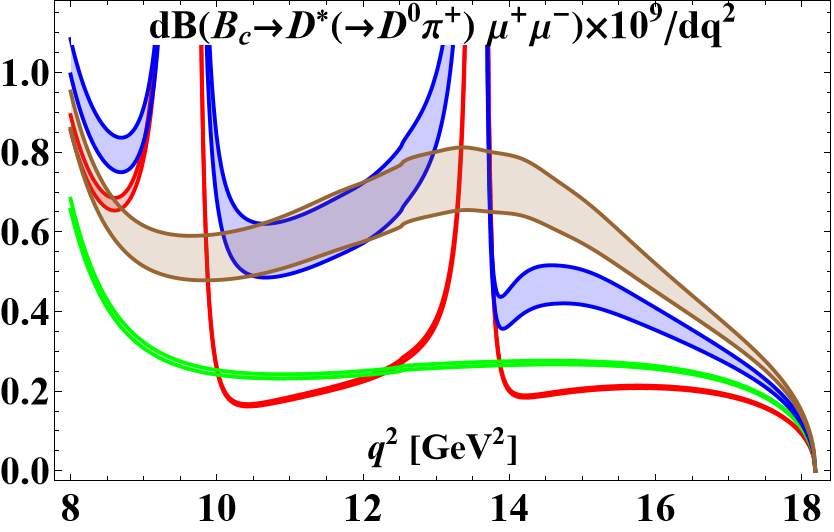}
\includegraphics[width=2.2in,height=1.36in]{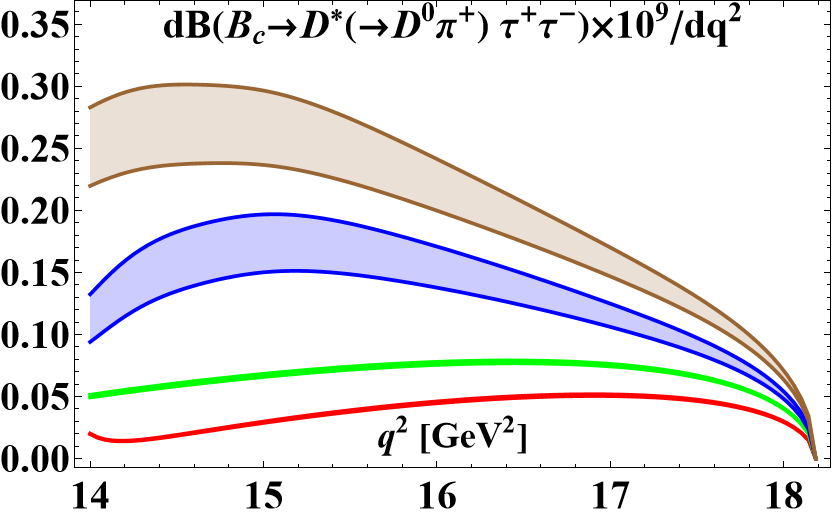}

\caption{Branching ratios of the two channels $B_c \to D^*(\to P_1 P_2)\ell^+\ell^- $ decays in the PB, PB+LD, PB+WA and PB+WA+LD contributions. The first and the second columns show the differential branching ratios $\mathrm{d}B/\mathrm{d}q^2$, with $\ell= \mu$, as a function of the squared dilepton mass $q^2$, in the lower and the higher $q^2$ regions, respectively, While the third column displays the differential branching ratios with final stage particle $\ell = \tau$, in the kinematical region $q^2=[15, q_{\text{max}}^2]~\text{GeV}^2$.}

\label{Fig1}
\end{figure}

The upper and lower plots in first column in FIG. \ref{Fig1} present the differential branching ratios for the decay channels $B_c \to D^*(\to D^+ \pi^0)\mu^+\mu^- $ and $B_c \to D^*(\to D^{0} \pi^{+})\mu^+\mu^- $, respectively, at low $q^2$ bin. Although light vector resonances such as $\rho$, $\omega$, and $\phi$ manifest as a small structure in the low-$q^2$ region, their overall impact on the decay spectrum is negligible, consequently, the green and red bands overlap. However, above $q^2 \gtrsim 4~\text{GeV}^2$, the red and green bands begin to separate due to the influence of the $J/\psi$ and $\psi(2S)$ resonances. The brown band represents the PB+WA contributions and is distinguishable from the green band over the entire $q^2$ range, demonstrating the significance of WA effects, particularly in the low-$q^2$ region. This sensitivity reflects the dependence of the branching ratio on the WA form factors as presented in the expressions of helicity amplitudes
Eqs \ref{helicityamps}-\ref{HA6}. Notably, the blue band, depicting the effects of LD contributions with PB and WA, where one can see, it overlaps with the brown band up to $7~\text{GeV}^2$. This suggests that, once all contributions are taken into account, the LD effects remain negligible up to $q^2 \sim 7~\text{GeV}^2$. Consequently, the theoretically reliable analysis window can be extended from the previously suggested range $q^2 \in [0.2, 5]~\text{GeV}^2$~\cite{Ju:2013oba} to $q^2 \in [0.2, 7]~\text{GeV}^2$. Therefore LD contributions minimally impact branching ratios and the other angular observables (as will be seen later in the analysis) at low $q^2$ region which make them suitable for NP parameters and can be reliably extracted within the $q^2$ range of $0.2$-$7~\text{GeV}^2$.

Similarly, the second column displays the behavior of $B_c \to D^*(\to D^{+,0} \pi^{0,+})\mu^+\mu^-$ decays in the high $q^2$ bin. The absence of overlap between the green and red bands across the entire region indicates substantial resonance contributions. It is well established in literature that at high $q^2$ region the charmonium resonances $J/\psi$ and $\psi(2S)$ begin dominating above $q^2 \approx 9~(\text{GeV}^2)$. In our case, these resonant states suppress the decay rate at high $q^2$ region as can be seen by the red band. While the WA contributions significantly enhance the decay amplitude. Notably, the brown and blue bands exhibit overlap exclusively within the $q^2$ interval $[10.5, 12.5]~\text{GeV}^2$ which indicates the resonance effects are small in this region, consequently, provide a clean window to measure the values of branching ratio in this region. However, in the region above $q^2\geq14~\text{GeV}^2$ the value of branching ratio is suppressed by inclusion of resonance effect as also seen in the work of Ju \textit{et al.} \cite{Ju:2013oba} but in the work of IPSS \cite{Ivanov:2024iat} we see slight increase in the branching ratio. This could be the artifact of obvious difference between form factors, also there is a difference in our works where we have used the low- and high-energy renormalization scale WCs separately.

Finally, the third column presents the branching ratios for $B_c \to D^*(\to D^{+,0} \pi^{0,+})\tau^+\tau^- $ decays. We have found that the behavior of the curves is qualitatively similar to that observed in the second column. As the red band lying systematically below the green band indicates significant suppression from resonance contributions. On the other hand, the WA contribution enhance the decay rate throughout the available $q^2$ region.

\subsubsection{Forward-backward asymmetry and Helicity fraction}

\begin{figure}[H]
\centering
\includegraphics[width=2.2in,height=1.36in]{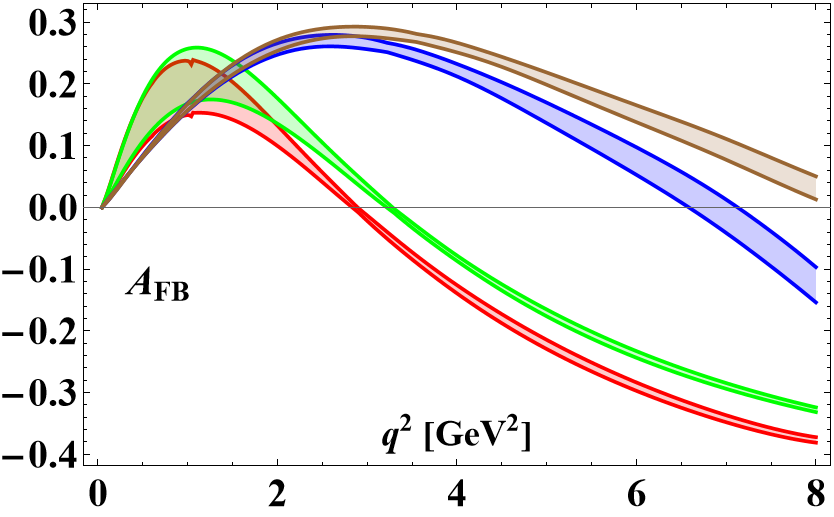}
\includegraphics[width=2.2in,height=1.36in]{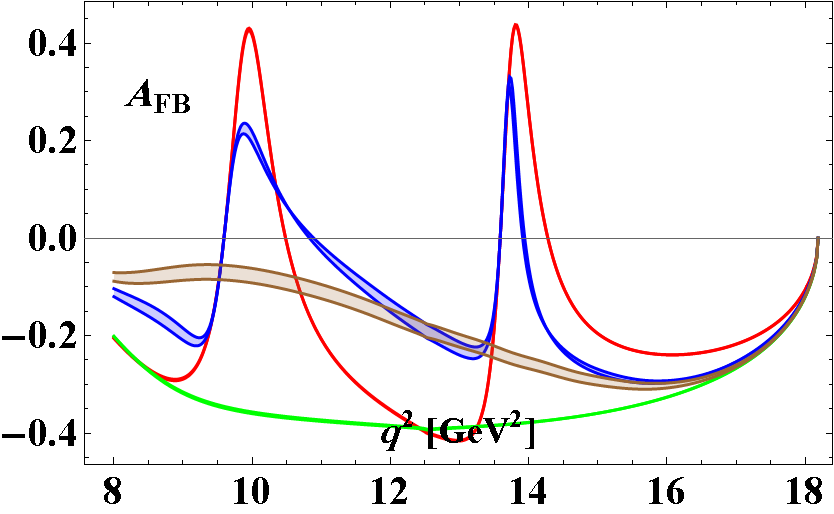}
\includegraphics[width=2.2in,height=1.36in]{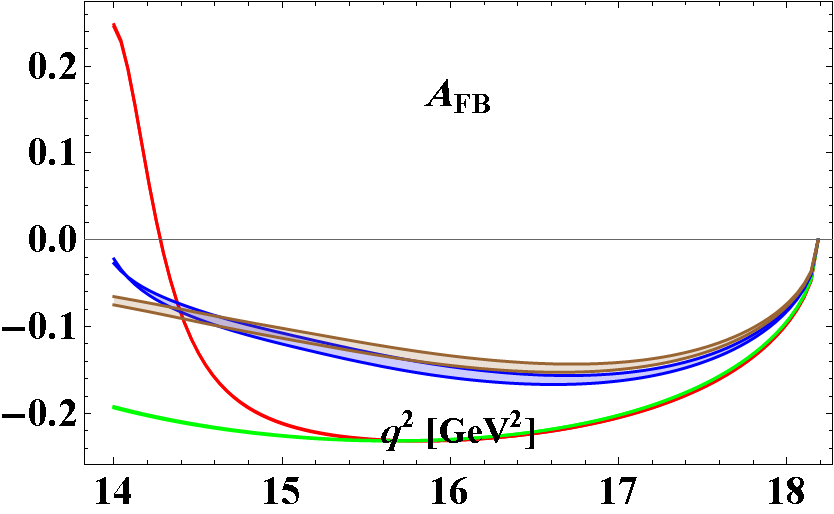}

\includegraphics[width=2.2in,height=1.36in]{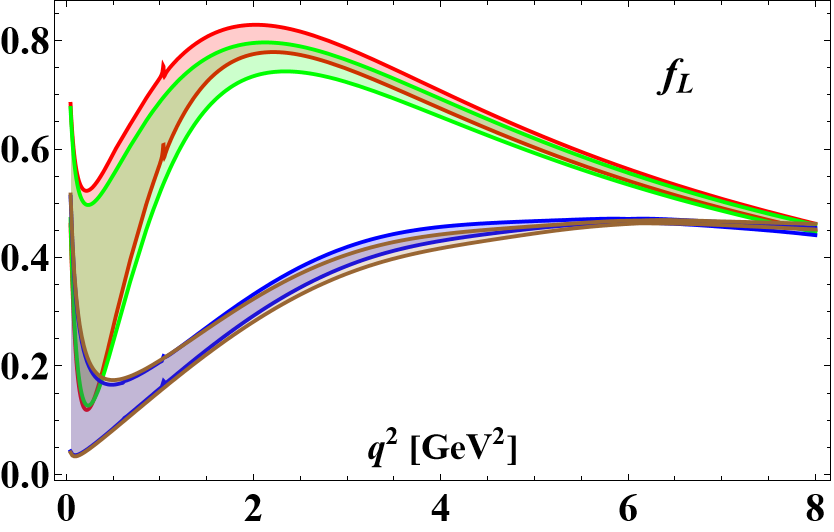}
\includegraphics[width=2.2in,height=1.36in]{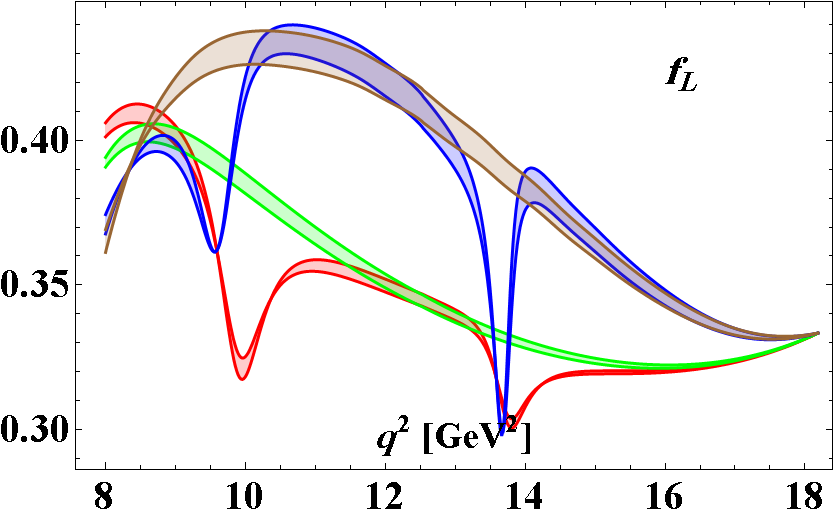}
\includegraphics[width=2.2in,height=1.36in]{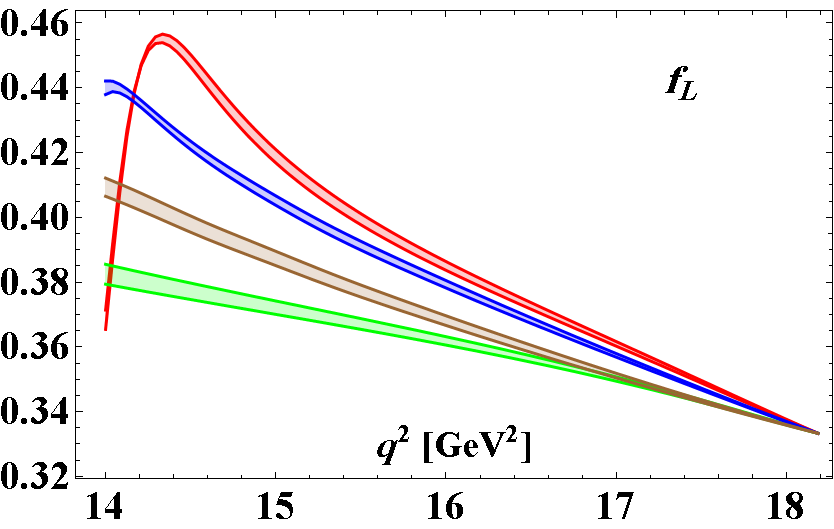}

\caption{The first and the second columns show (from top to bottom) the $A_{\mathrm{FB}}$ and $f_L$ in the PB, PB+LD, PB+WA and PB+WA+LD contributions in the lower and the higher  for $\ell = \mu$, as functions of the squared dilepton mass $q^2$ $(\text{GeV}^2)$ respectively. While the third column displays the same observables with final stage particles $\ell = \tau$ in the higher bin only. Legends are similar as described in FIG.~\ref{Fig2}.}
\label{Fig2}
\end{figure}

FIG. \ref{Fig2} depicts the behavior of the forward-backward asymmetry $A_{FB}$ (1st row) and the longitudinal helicity fraction $f_L$ (2nd row) as a function of $q^{2}$. The first two columns in FIG. \ref{Fig2} show the behavior of the said observables for the decay $B_{c}\to D^{\ast}\mu^{+}\mu^{-}$ in low and high $q^2$ bin, while the same observables are presented in the third column for the $B_{c}\to D^{\ast}\tau^{+}\tau^{-}$ decay. By the comparison of green and red bands it can be observed that the zero crossing of $A_{FB}(B_{c}\to D^{\ast}\mu^{+}\mu^{-})$ shifts slightly to the left due to LD contributions. This slight shift of zero crossing is also visible in the work of IPSS \cite{Ivanov:2024iat}. Similar to the branching ratio, the WA enhances the amplitude of the $A_{FB}$ which can also be seen in the work of Ju \textit{et al} \cite{Ju:2013oba}. In addition, the zero crossing of $A_{FB}$ shifts significantly above the $\gtrsim8$ GeV$^2$ when the WA effects are incorporated. However, when we include all the contributions, the zero crossing lying at around $\simeq7$ GeV$^2$ as can be seen by the blue band. One can also see the effects of LD at low $q^2$ are also mild for the $A_{FB}$, particularly, below the $q^{2}\simeq3$ GeV$^2$ region.

The second plot in first row illustrates $A_{FB}$ behavior in the high-$q^2$ region. At $q^2\gtrsim 15~\text{GeV}^2$, the resonance effects are negligible that can be seen  by overlapping between the blue and brown bands. Therefore, this region is clean and suitable for measurements. Finally in the last plot of the first row we see that brown and blue bands overlap, that also suggests the clean region above $q^2\approx 15~\text{GeV}^2$. Consequently, we observe that the $A_{FB}$ serves as a good observable to test any NP effects in the high-$q^2$ window $[15, q_{\text{max}}^2]~\text{GeV}^2$.

The first plot in the second row of FIG.~\ref{Fig1} displays the $q^2$ dependence of the  $f_{L}$ in the low-$q^2$ region. The substantial overlap between green and red bands, as well as between brown and blue bands, indicates that LD contributions have almost negligible impact on this observable throughout this low $q^2$ region. The second plot reveals that the clean regions are in the bins $\sim[10, 13]~\text{GeV}^2$ and $\sim[14, q_{\text{max}}^2]~\text{GeV}^2$ which can also be seen by overlapping between the blue and brown bands. The third plot of the second row corresponds to the case of $\tau$'s as final state leptons. The maximum difference in the blue and brown bands is at $q^2\simeq14$GeV$^2$ which is about less than 3\%. Therefore, in the case of $\tau$'s, the value of $A_{FB}$ is mildly effected by the resonances. In addition, the thin bands for each contribution indicates that the uncertainties of different form factors almost cancel each other and hence it leads to relatively clean observable for comparison with the precise experimental predictions.

\subsubsection{Phenomenological analysis of Angular coefficients in 
\texorpdfstring{$B_{c}\to D^{\ast}\ell^{+}\ell^{-}$}
{Bc+ -> D*+ l+ l-} Decays}

\begin{figure}[h]
\centering

\includegraphics[width=2.2in,height=1.36in]{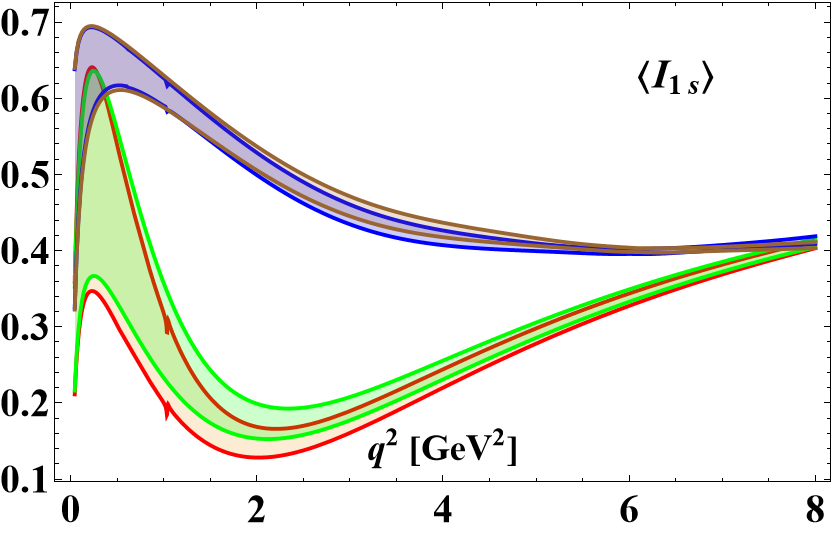}
\includegraphics[width=2.2in,height=1.36in]{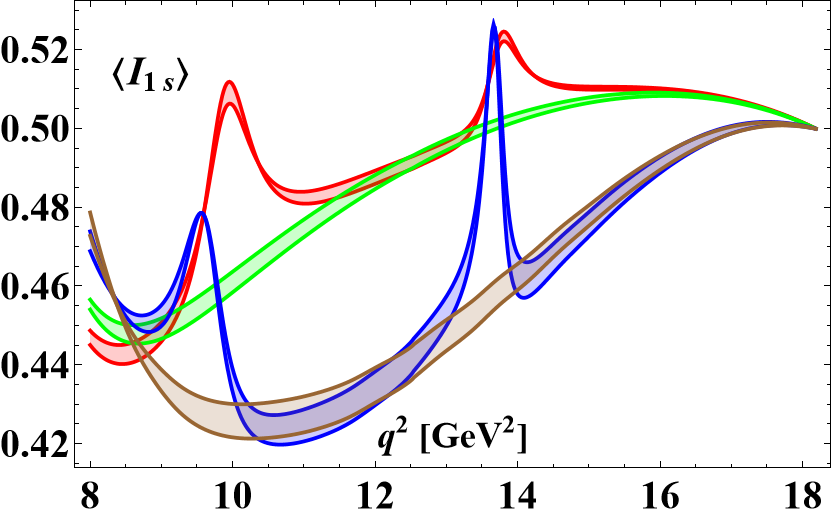}
\includegraphics[width=2.2in,height=1.36in]{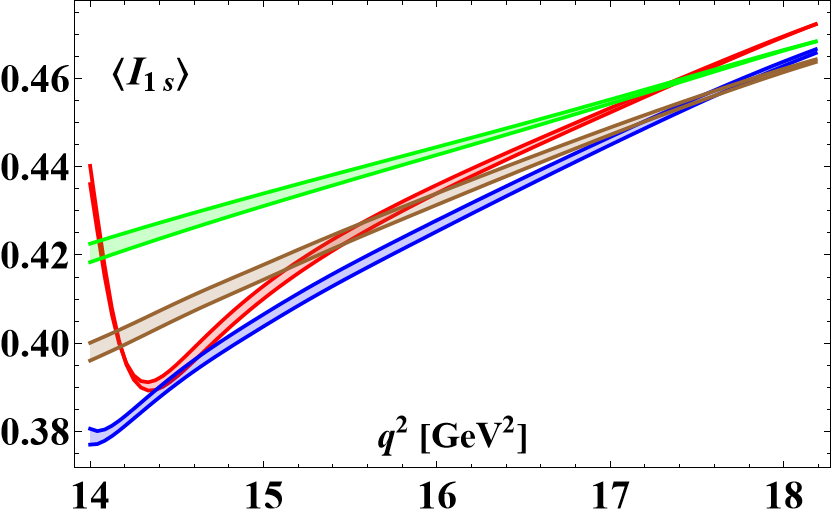}

\includegraphics[width=2.2in,height=1.36in]{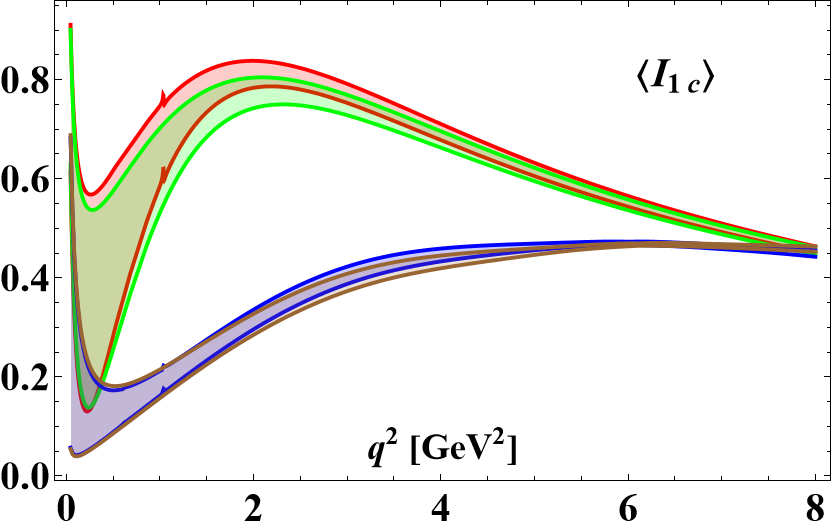}
\includegraphics[width=2.2in,height=1.36in]{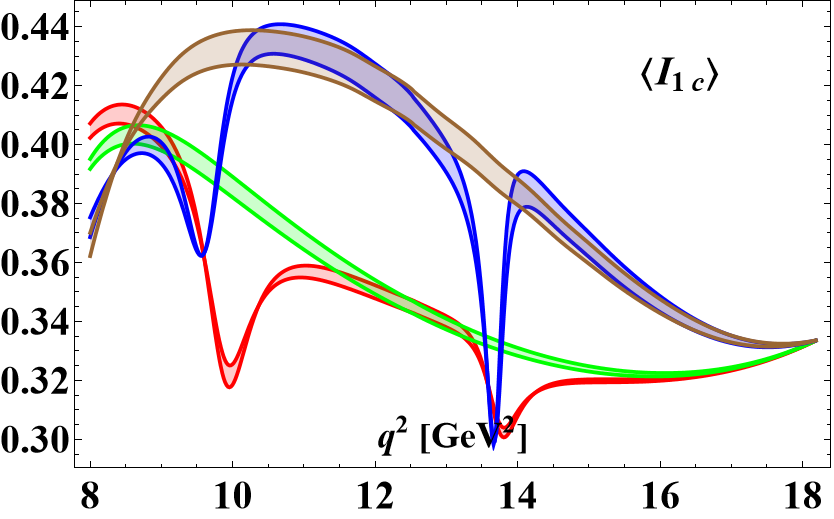}
\includegraphics[width=2.2in,height=1.36in]{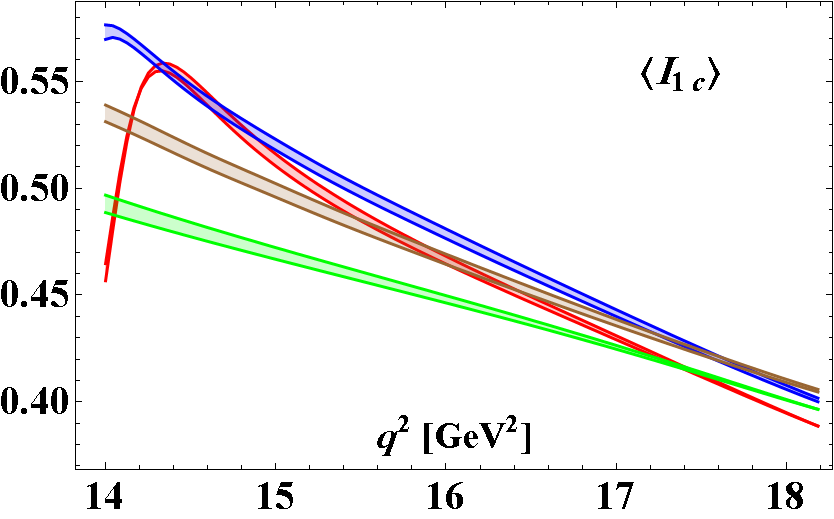}

\includegraphics[width=2.2in,height=1.36in]{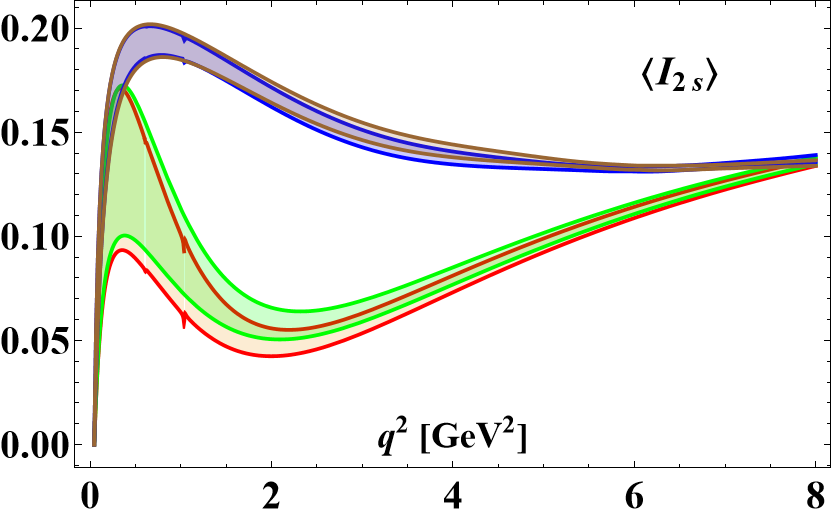}
\includegraphics[width=2.2in,height=1.36in]{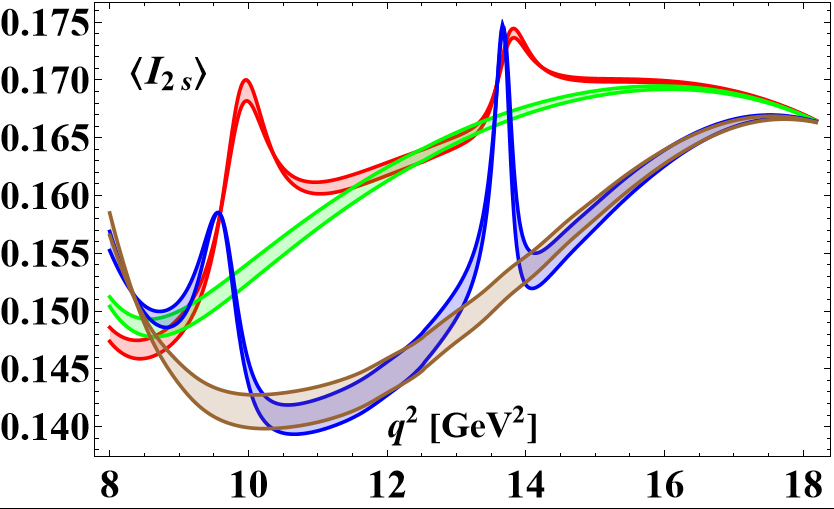}
\includegraphics[width=2.2in,height=1.36in]{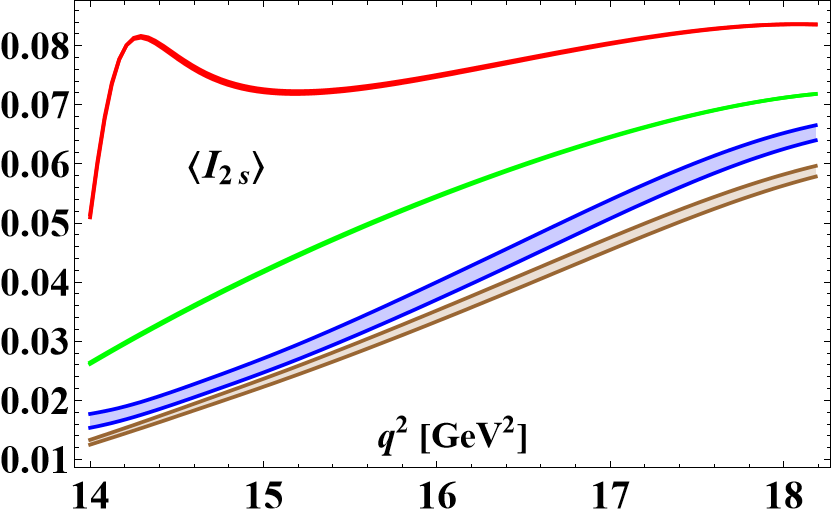}

\includegraphics[width=2.2in,height=1.36in]{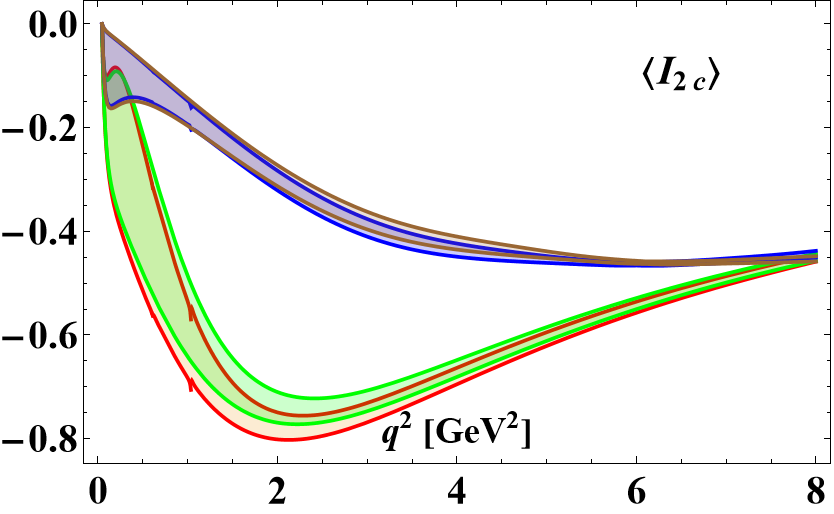}
\includegraphics[width=2.2in,height=1.36in]{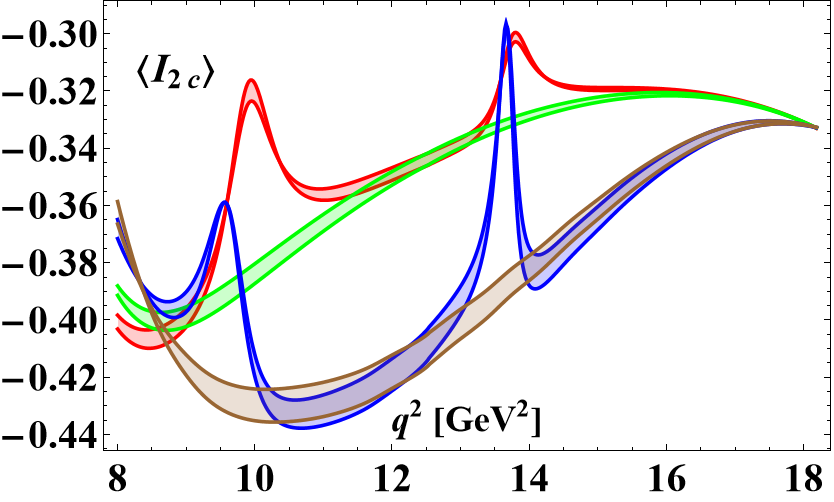}
\includegraphics[width=2.2in,height=1.36in]{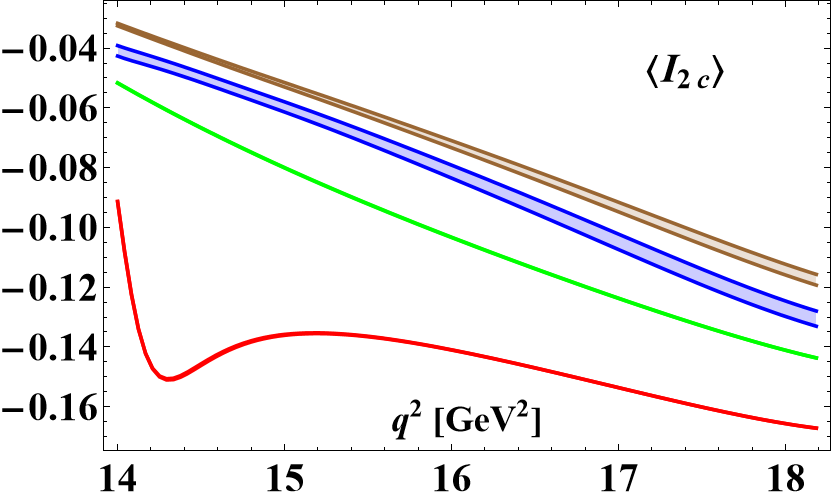}

\caption{Angular observables of $B_c \to D^*(\to P_1 P_2)\ell^+\ell^- $ decay in the PB, PB+LD, PB+WA and PB+WA+LD contributions. The first and the second columns show the average values of angular observables $\langle I_{(1s,1c,2s,2c)} \rangle$ in the lower and the upper bins for $\ell= \mu$ as functions of the squared dilepton mass $q^2$ $(\text{GeV}^2)$, respectively. While the third column displays the same observables with final stage particles $\ell= \tau$ in the higher bin.}
\label{Fig3}
\end{figure}

\begin{figure}[H]
\centering

\includegraphics[width=2.2in,height=1.36in]{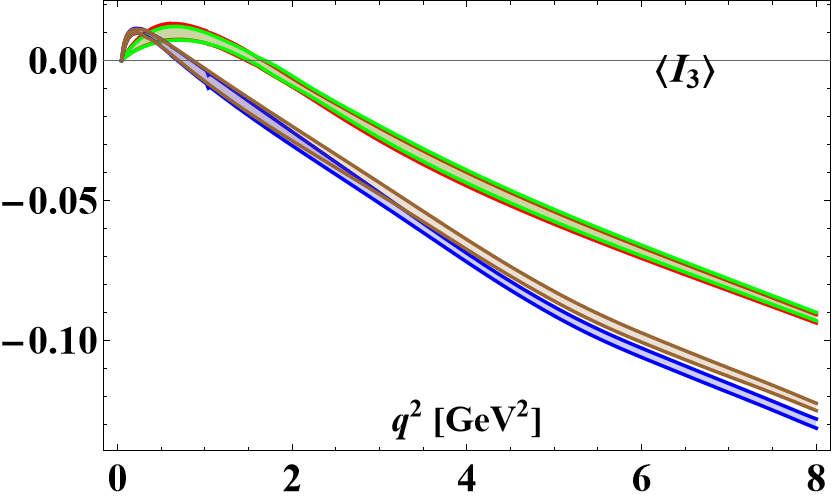}
\includegraphics[width=2.2in,height=1.36in]{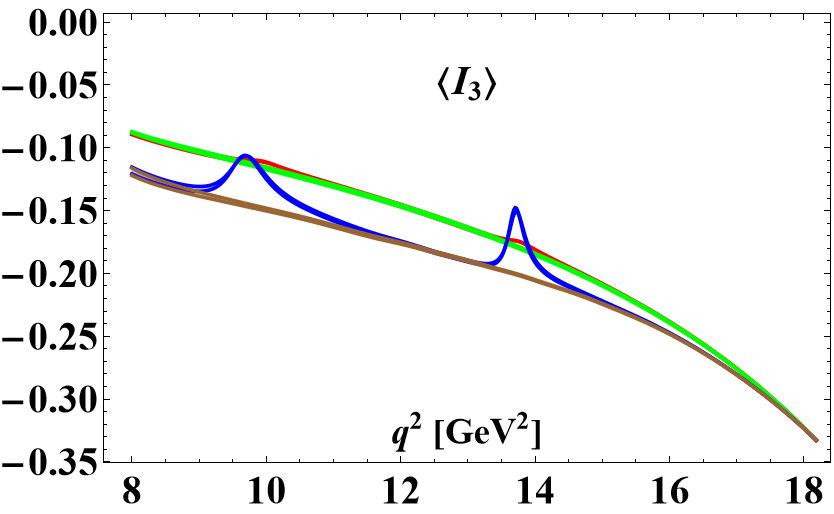}
\includegraphics[width=2.2in,height=1.36in]{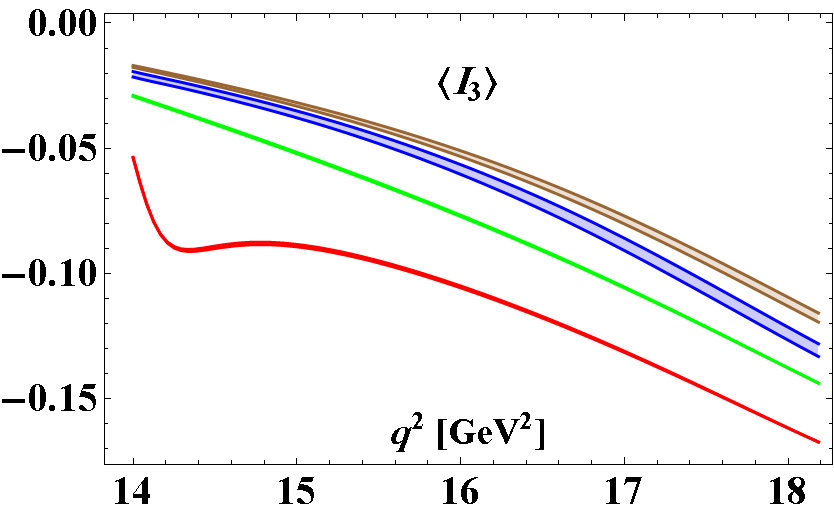}

\includegraphics[width=2.2in,height=1.36in]{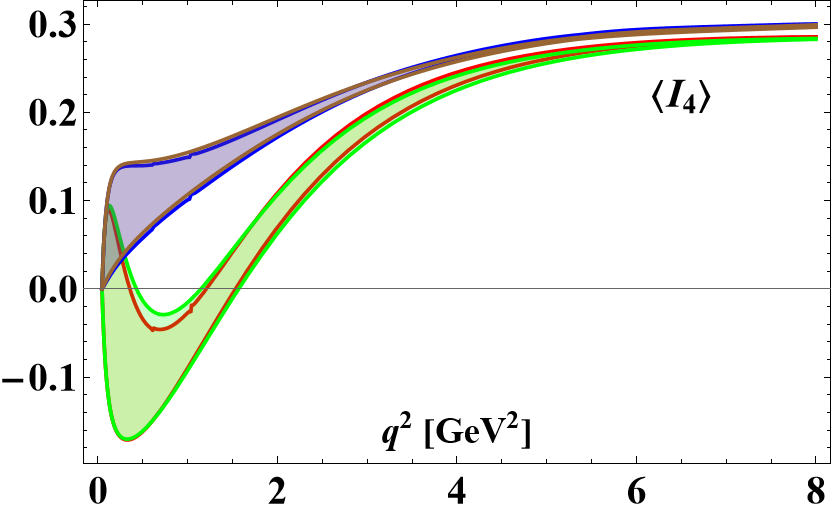}
\includegraphics[width=2.2in,height=1.36in]{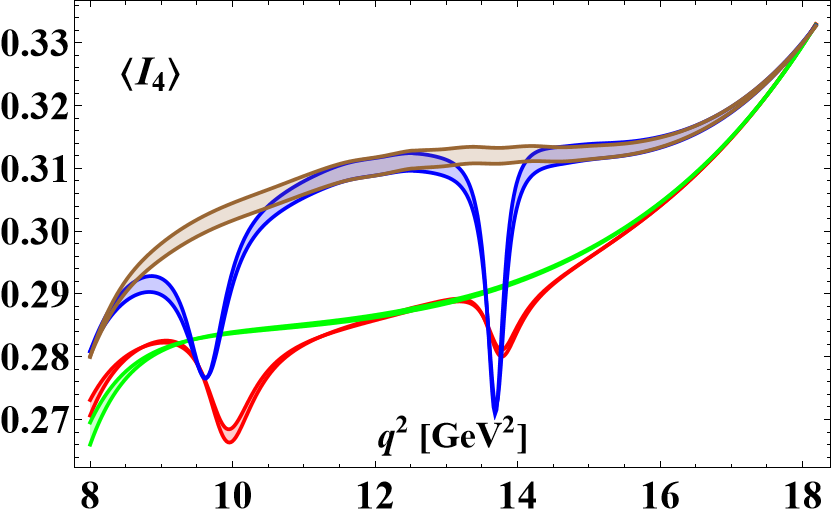}
\includegraphics[width=2.2in,height=1.36in]{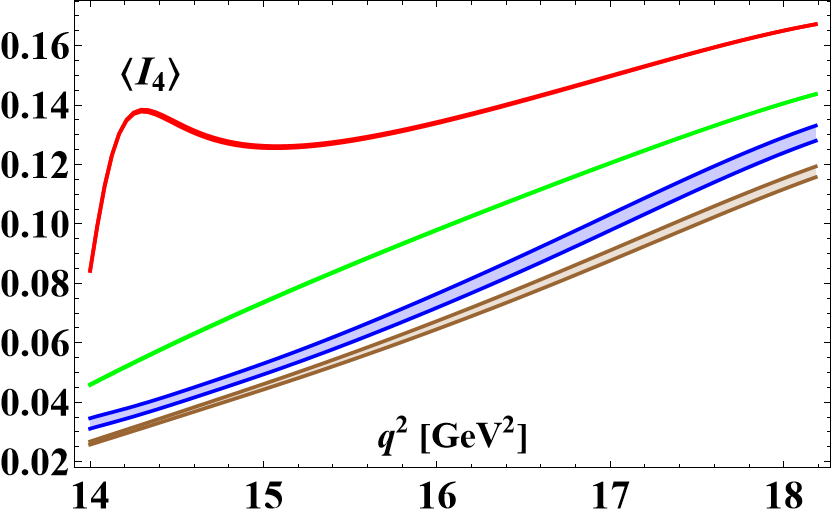}

\includegraphics[width=2.2in,height=1.36in]{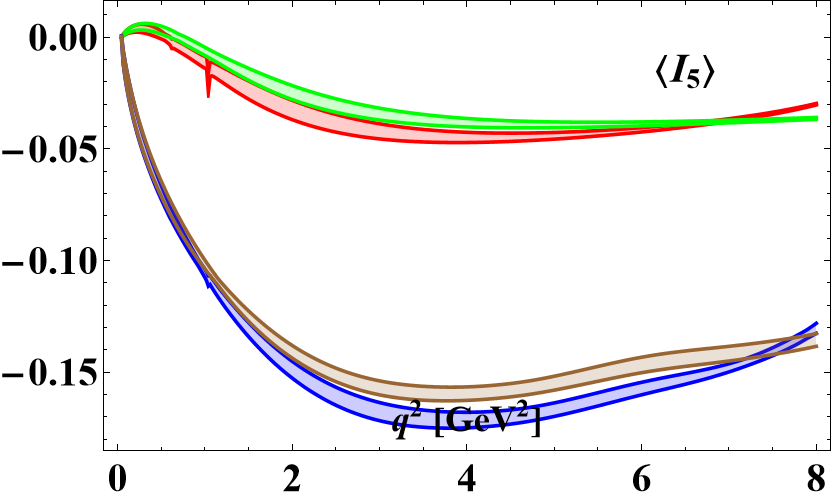}
\includegraphics[width=2.2in,height=1.36in]{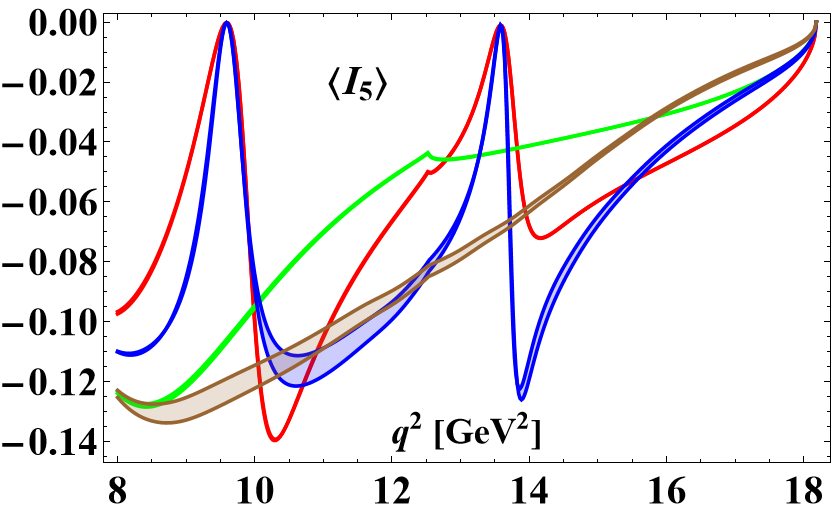}
\includegraphics[width=2.2in,height=1.36in]{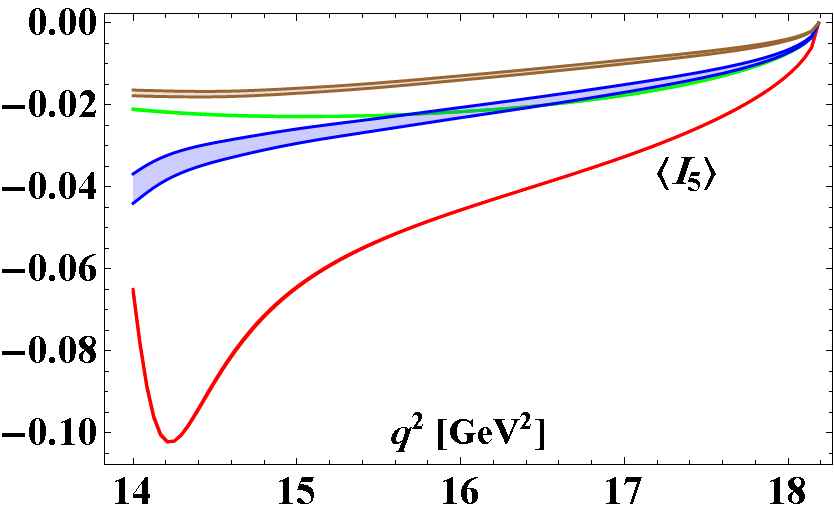}

\includegraphics[width=2.2in,height=1.36in]{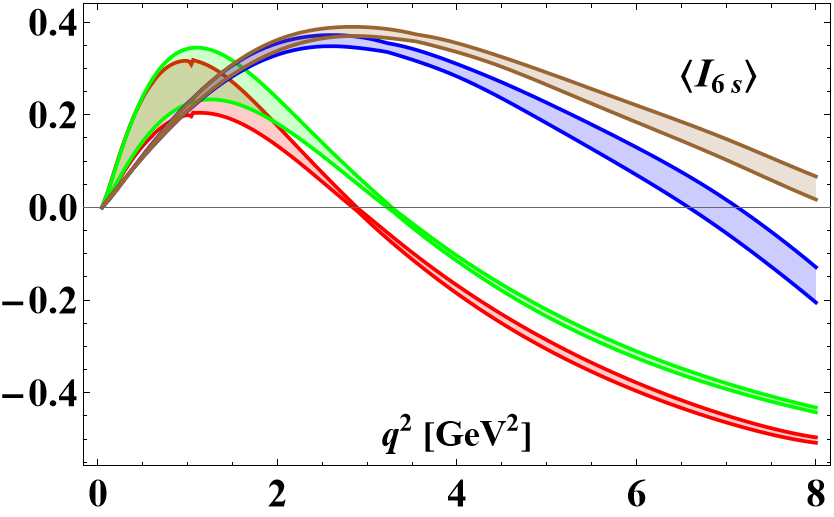}
\includegraphics[width=2.2in,height=1.36in]{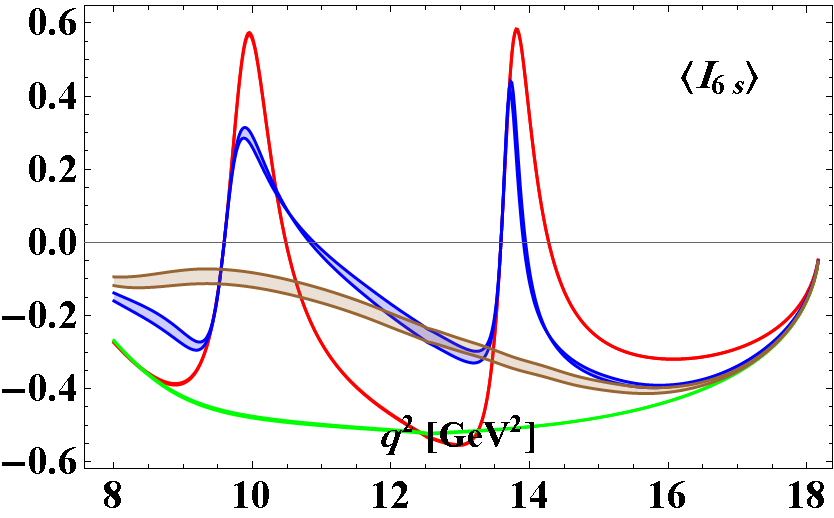}
\includegraphics[width=2.2in,height=1.36in]{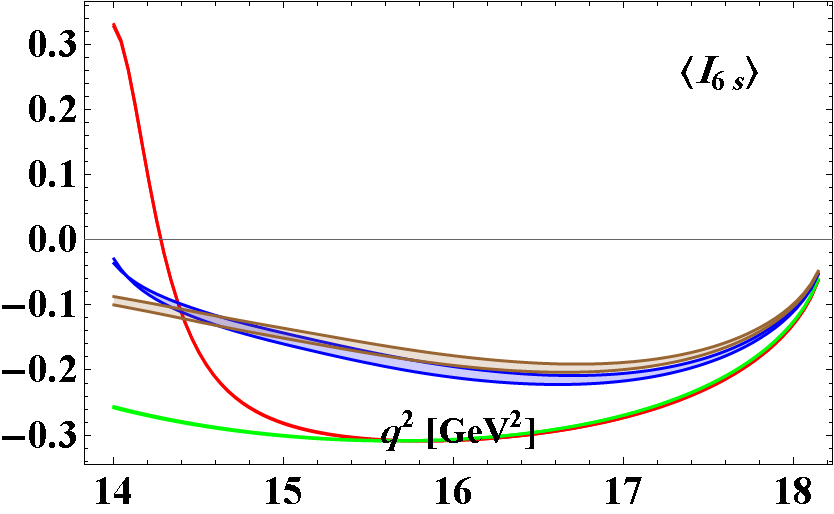}

\caption{Angular observables of $B_c \to D^*(\to P_1 P_2)\ell^+\ell^- $ decay in the PB, PB+LD, PB+WA and PB+WA+LD contributions. The first and the second columns show the average values of angular observables $\langle I_i \rangle$ in the lower and the upper bins for $\ell = \mu$ as functions of the squared dilepton mass $q^2$ $(\text{GeV}^2)$, respectively. While the third column displays the same observables with final stage particles $\ell = \tau$ in the higher bin only.}
\label{Fig4}
\end{figure}

In this section, we present the WA effects in the individual angular coefficients $\langle I_{1s}\rangle,
\langle I_{1c}\rangle,
\langle I_{2s}\rangle,
\langle I_{2c}\rangle,
\langle I_{3}\rangle,
\langle I_{4}\rangle,
\langle I_{5}\rangle$ and $\langle I_{6s}\rangle$. Numerical values of the SM predictions of the average angular coefficients with uncertainties due to the form factors in different $q^{2}$ are presented in Table-\ref{binned}. Furthermore, the plots of the angular coefficients as a function of $q^{2}$ are presented in FIG.~\ref{Fig3} and FIG. \ref{Fig4}.

FIG.~\ref{Fig3} presents the angular observables $\langle I_{1s}\rangle,
\langle I_{1c}\rangle,
\langle I_{2s}\rangle$ and $
\langle I_{2c}\rangle$, with the first column displaying their behavior in the low-$q^2$ region and the second column showing the high-$q^2$ domain for $\mu$ as the final meson state. The third column depicts the same observables for the $\tau$ as final state meson, at high $q^2$. In the low-$q^2$ region for the case of $\mu$, the substantial overlap between green and red bands, as well as between brown and blue bands, indicates $\rho,\omega$ and $\phi$ give negligible contributions to these observables. The tail effects of $J/\psi$ and $\psi(2S)$ are also absent in the low $q^2$ bin for the said angular observables. Consequently, LD effects can be safely neglected when studying NP models in this kinematic regime. This finding suggests the theoretically clean window to $q^2\in [0.045, 8]~\text{GeV}^2$ for these angular observables. However, WA contributions exhibit significant enhancement across all observables in the $q^2$ range $[0.5, 7]~\text{GeV}^2$ (which can be seen by green and brown bands), demonstrating the dominant role of this mechanism in the low-$q^2$ region. 

Similarly in the second column, the pronounced separation between the brown and green bands also shows substantial WA contributions across the high-$q^2$ region. The green and red bands exhibit limited overlap only in the narrow interval between resonance peaks and above $15~\text{GeV}^2$. However, when we include all the contributions, the resonance free region is in the intervals $[10, 13]~\text{GeV}^2$ and $[14, q_{\text{max}}^2]~\text{GeV}^2$ which can easily be seen by the overlap between the brown and blue bands.

The third column displays qualitatively similar trends, though with enhanced deviation between red and green bands across all observables. The vertical scale indicates minimal variation (less than 5\%) between brown and blue bands throughout $q^2$ region. For $\langle I_{2s} \rangle$ and $\langle I_{2c} \rangle$, these bands remain closely aligned with each other across the entire $q^2$ range, while for $\langle I_{1s} \rangle$ and $\langle I_{1c} \rangle$, a little larger separation occurs below $15~\text{GeV}^2$. These observations not only reaffirm the previously identified theoretically clean window of $[15, q_{\text{max}}^2]~\text{GeV}^2$ ~\cite{Ju:2013oba}, but also suggest a possible extension to $[14, q_{\text{max}}^2]~\text{GeV}^2$.

FIG.~\ref{Fig4} shows the remaining angular observables 
$\langle I_{3} \rangle$, $\langle I_{4} \rangle$, 
$\langle I_{5} \rangle$, and $\langle I_{6s} \rangle$. 
While the overall behavior is similar to that observed in Fig.~\ref{Fig3}, 
notable differences appear for $\langle I_{3} \rangle$, 
$\langle I_{5} \rangle$, and, in particular, $\langle I_{6s} \rangle$. 
In the low-$q^2$ region, $\langle I_{3} \rangle$ and 
$\langle I_{5} \rangle$ exhibit mild separations between the brown and blue bands, 
emerging around $q^2 \sim 3~\text{GeV}^2$ and 
$q^2 \sim 2~\text{GeV}^2$, respectively. As a consequence, the clean region that is minimally affected by resonances 
is reduced to lower values of $q^2$ for both observables. On the other hand the $A_{FB}\propto\langle I_{6s} \rangle$ (see Eq. \ref{AFB1}) hence follow the same trend as forward-backward asymmetry in the low $q^2$ bin as discussed earlier.

For $\langle I_{4} \rangle$, the inclusion of the WA contribution (brown and blue bands) constrains this observable to remain strictly positive, whereas the PB contributions alone (red and green bands) yield negative values for $q^2 \lesssim 2~\text{GeV}^2$. In a similar manner, $\langle I_{5} \rangle$ exhibits marginally positive values in the PB-only contribution, while the inclusion of WA contributions renders it entirely negative.

In the high-$q^2$ region, $\langle I_{5} \rangle$ exhibits a pronounced sensitivity to resonance effects compared to the other angular observables. By contrast, the brown and blue bands for $\langle I_{3} \rangle$ and $\langle I_{4} \rangle$ largely overlap within the intervals $q^2 \in [10.5, 13]~\text{GeV}^2$ and $q^2 \in [14, q_{\text{max}}^2]$, respectively, whereas $\langle I_{6s} \rangle$ shows comparable agreement only in the more restricted region $q^2 \in [15, q_{\text{max}}^2]$.

The third column, corresponding to the $\tau$ channel, exhibits only a mild separation between the brown and blue bands, with their relative difference remaining below $5\%$. This indicates that these observables in the $\tau$ mode are largely insensitive to resonance effects and therefore provide a clean probe for comparisons between experimental measurements and SM as well as beyond the SM predictions.

In summary, the observables $\langle I_{3} \rangle$, $\langle I_{4} \rangle$, and $\langle I_{5} \rangle$ provide reliable probes in the low-$q^2$ region, in contrast to $\langle I_{6s} \rangle$. In the high-$q^2$ domain, $\langle I_{3} \rangle$ and $\langle I_{4} \rangle$ remain theoretically clean within the intervals $q^2 \in [10.5, 13]~\text{GeV}^2$ and $q^2 \in [14, q_{\text{max}}^2]$, respectively, whereas $\langle I_{6s} \rangle$ is reliable only in the more restricted region $q^2 \in [15, q_{\text{max}}^2]$. By contrast, $\langle I_{5} \rangle$ is less suitable for high-$q^2$ analyses due to sizable resonance-induced effects.

\begin{figure}[H]

\centering
\includegraphics[width=2.2in,height=1.36in]{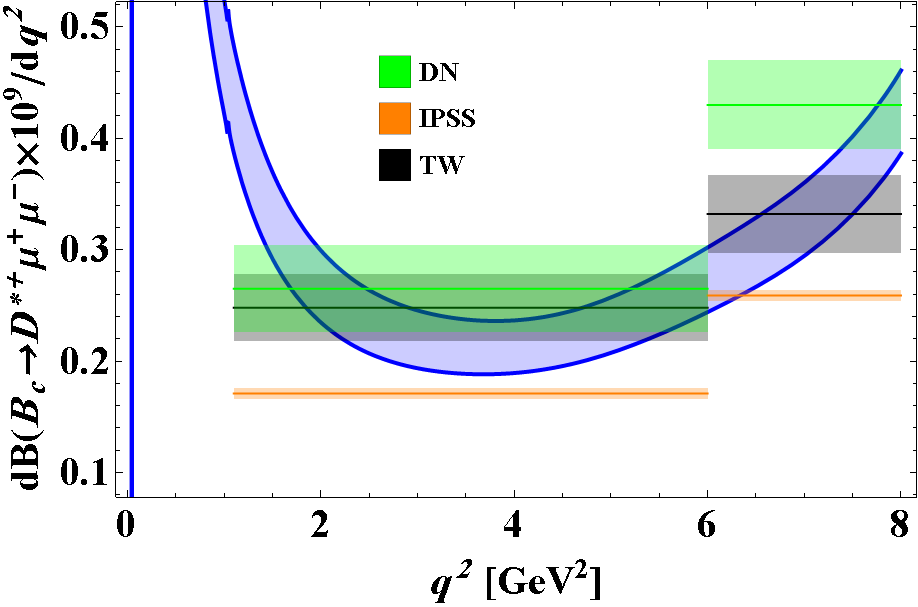}
\includegraphics[width=2.2in,height=1.36in]{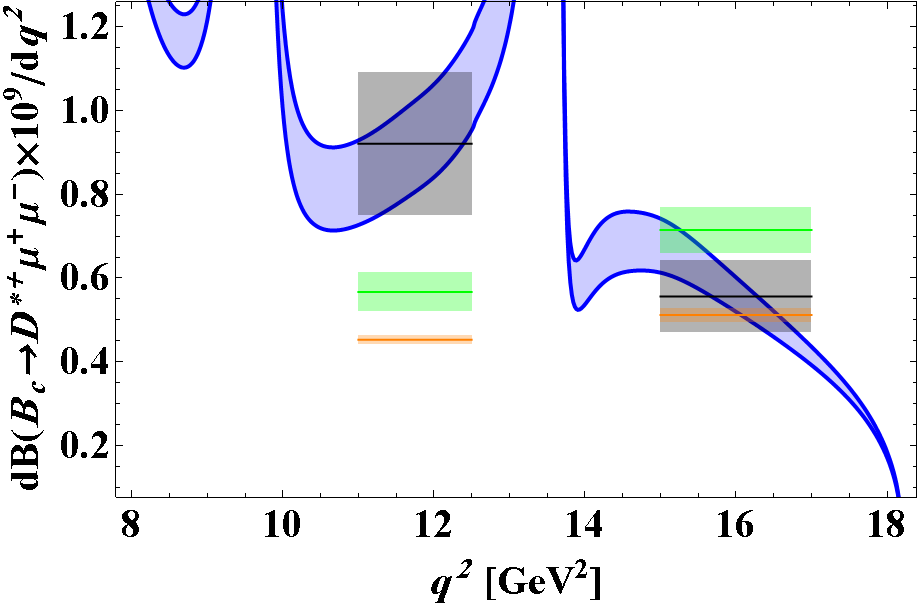}
\includegraphics[width=2.2in,height=1.36in]{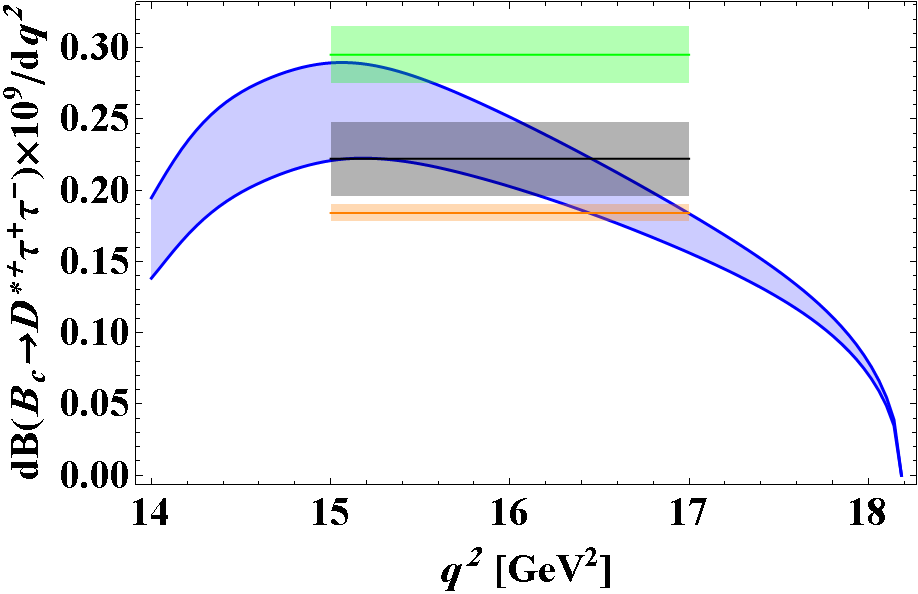}

\includegraphics[width=2.2in,height=1.36in]{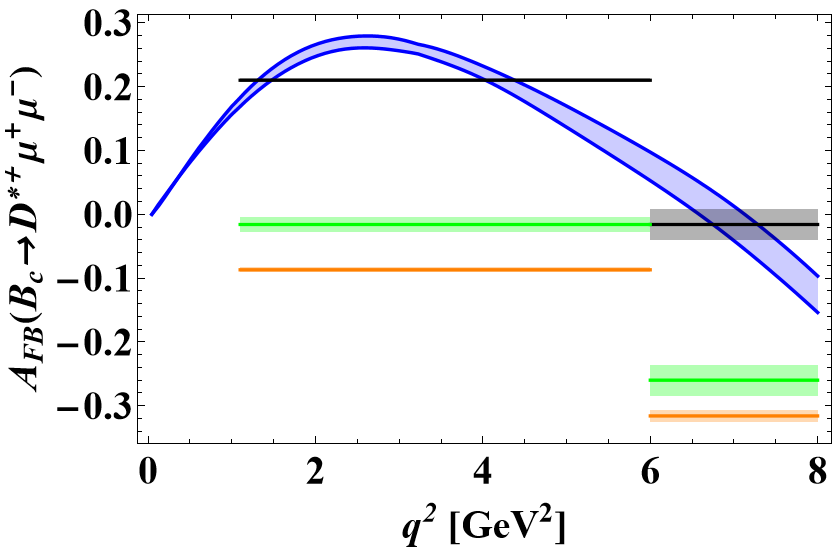}
\includegraphics[width=2.2in,height=1.36in]{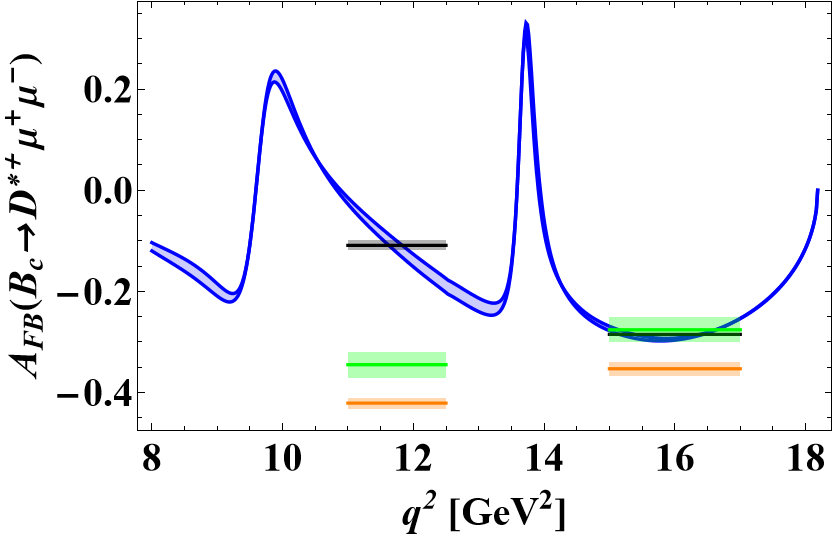}
\includegraphics[width=2.2in,height=1.36in]{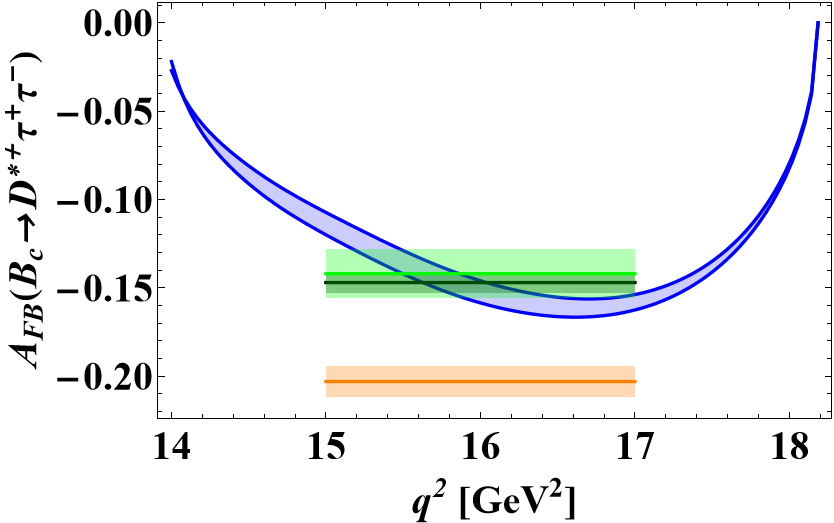}

\includegraphics[width=2.2in,height=1.36in]{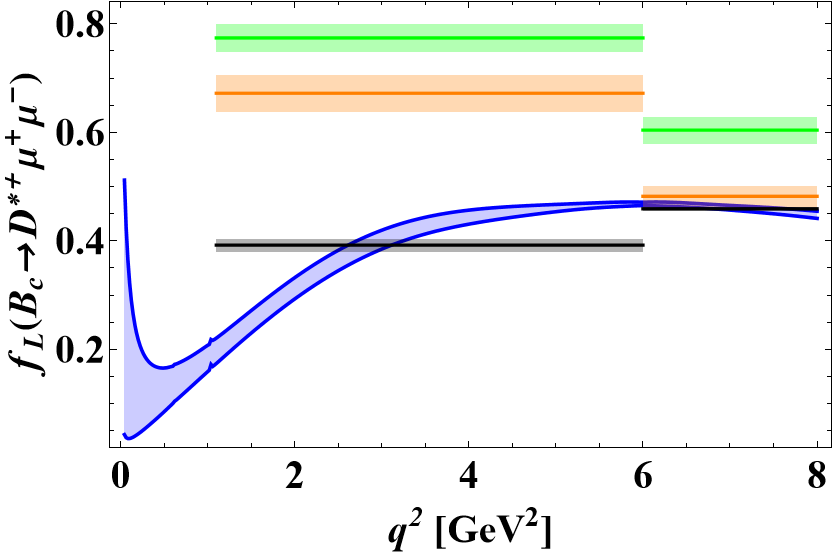}
\includegraphics[width=2.2in,height=1.36in]{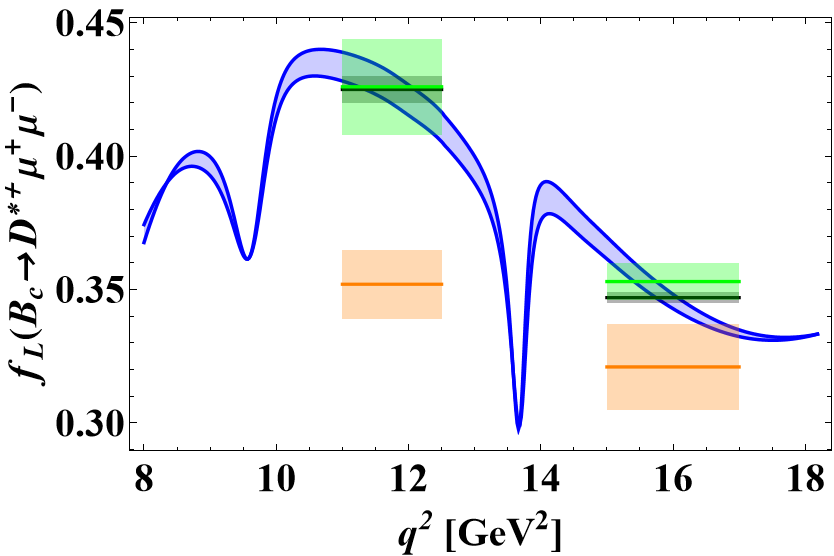}
\includegraphics[width=2.2in,height=1.36in]{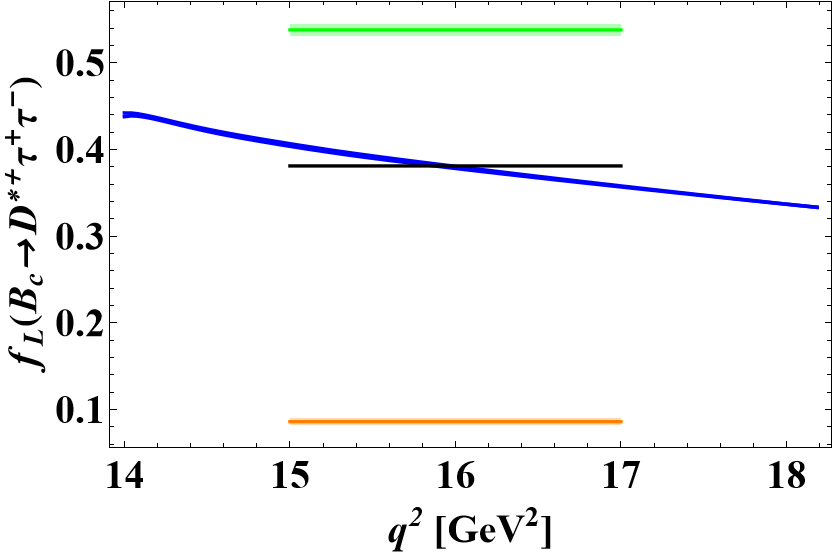}
\caption{A comparison of the binned average values of the branching ratio, forward backward asymmetry and helicity fraction as provided by IPSS~\cite{Ivanov:2024iat} (orange), DN~\cite{Dey:2025xjg} (green) and this work (TW, black).}
\label{comparison}
\end{figure}

In FIG.~\ref{comparison}, we present a comparison of binned average values obtained by IPSS \cite{Ivanov:2024iat} using form factors calculated in the CCQM (orange bands), by DN \cite{Dey:2025xjg} using form factors calculated in pQCD (green bands), and the results of this work (TW), which use the IPSS~\cite{Ivanov:2024iat} PB form factors while also incorporating the WA contributions (black bands). It should be emphasized that the other two works include only the PB contributions, whereas in our case we have also included the WA effects. Since our analysis employs the PB form factors calculated by IPSS~\cite{Ivanov:2024iat}, a direct comparison between the black and orange bands is meaningful to isolate the effects of the WA contributions on the considered observables. The comparison between the green and orange bands illustrates the sensitivity of the observables to the choice of form factors and highlights that a precise determination of form factors is essential before probing potential NP effects.

The black bands in the branching ratio plots (first row) consistently lie above the corresponding orange bands. In particular, a pronounced separation between the two appears in the $[11,\,12.5]~\text{GeV}^2$ bin, as shown in the second plot of the first row. In the second row, sizable deviations between the black and orange bands occur across all bins of $A_{FB}$, demonstrating the strong sensitivity of this observable to WA effects. A similarly significant separation is visible in the predictions for $f_L$. Overall, this comparison provides a clear and comprehensive illustration of the impact of WA contributions on the relevant observables. Finally we present our prediction of the binned average values of the observables in Table~\ref{binned}.

The black bands in the branching-ratio plots (first row) consistently lie above the corresponding orange bands. In particular, a pronounced separation between the two is observed in the $[11,,12.5]~\text{GeV}^2$ bin, as shown in the second plot of the first row. In the second row, significant deviations between the black and orange bands appear across all bins of $A_{FB}$, highlighting the strong sensitivity of this observable to WA effects. A similarly notable separation is visible in the predictions for $f_L$. Overall, this comparison provides a clear and comprehensive illustration of the impact of WA contributions on the relevant observables. Finally, our predictions for the binned average values of all observables are summarized in Table \ref{binned}.

\begin{table}[H]
\centering
\renewcommand{\arraystretch}{1.25}
\begin{tabular}{|l|c|c|c|c|c|}
\hline
Observable & $\ell=\mu$ & $\ell=\mu$ & $\ell=\mu$ & $\ell=\mu$ & $\ell=\tau$  \\ & $[1.1-6]~\text{GeV}^2$ & $[6-8]~\text{GeV}^2$ & $[11-12.5]~\text{GeV}^2$ & $[15-17]~\text{GeV}^2$ & $[15-17]~\text{GeV}^2$  \\
\hline
$\mathrm{Br} \cross 10^{9}$ &
$0.248 \pm 0.030$ &
$0.332 \pm 0.035$ &
$0.921 \pm 0.170$ &
$0.556 \pm 0.086$ &
$0.222 \pm 0.026$ \\
\hline
$A_{\mathrm{FB}}$ &
$0.210 \pm 0.003$ &
$-0.016 \pm 0.024$ &
$-0.109 \pm 0.010$ &
$-0.284 \pm 0.002$ &
$-0.218 \pm 0.007$ \\
\hline
$f_L$ &
$0.392 \pm 0.012$ &
$0.459 \pm 0.004$ &
$0.425 \pm 0.005$ &
$0.347 \pm 0.002$ &
$0.585 \pm 0.001$ \\
\hline
$\langle I_{1s}\rangle$ &
$0.455 \pm 0.009$ &
$0.405 \pm 0.003$ &
$0.432 \pm 0.004$ &
$0.490 \pm 0.002$ &
$0.427 \pm 0.001$ \\
\hline
$\langle I_{1c}\rangle$ &
$0.464 \pm 0.082$ &
$0.461 \pm 0.067$ &
$0.426 \pm 0.005$ &
$0.348 \pm 0.003$ &
$0.480 \pm 0.002$ \\
\hline
$\langle I_{2s}\rangle$ &
$0.149 \pm 0.003$ &
$0.134 \pm 0.001$ &
$0.143 \pm 0.001$ &
$0.163 \pm 0.001$ &
$0.039\pm 0.001$ \\
\hline
$\langle I_{2c}\rangle$ &
$-0.387 \pm 0.012$ &
$-0.457 \pm 0.003$ &
$-0.423 \pm 0.005$ &
$-0.346 \pm 0.002$ &
$-0.082\pm 0.002$ \\
\hline
$\langle I_3\rangle$&
$-0.059 \pm 0.002$ &
$-0.117 \pm 0.002$ &
$-0.171 \pm 0.001$ &
$-0.249 \pm 0.000$ &
$-0.060\pm 0.002$ \\
\hline
$\langle I_4\rangle$ &
$0.236 \pm 0.005$ &
$0.296 \pm 0.001$ &
$0.310 \pm 0.001$ &
$0.315 \pm 0.001$ &
$0.075\pm 0.002$ \\
\hline
$\langle I_5\rangle$ &
$-0.160 \pm 0.004$ &
$-0.148 \pm 0.002$ &
$-0.100 \pm 0.003$ &
$-0.043 \pm 0.001$ &
$-0.022\pm 0.001$ \\
\hline
$\langle I_{6s}\rangle$ &
$0.279 \pm 0.004$ &
$-0.021 \pm 0.032$ &
$-0.145 \pm 0.014$ &
$-0.380 \pm 0.003$ &
$-0.196\pm 0.008$ \\
\hline
\end{tabular}
\caption{Binned average observables with uncertainties.}
\label{binned}
\end{table}

 \section{Conclusion}

The numerical analysis presented in this work provides a comprehensive assessment of the SM contributions to the decay $B_c \to D^{*}\ell^{+}\ell^{-}$. Our results demonstrate that WA plays a significant role in shaping the behavior of several observables, particularly in the low-$q^{2}$ region, where its interference with the PB amplitudes leads to noticeable modifications in the values of the observables. In contrast, LD effects are generally small in this region. In the high-$q^{2}$ domain, resonance contributions become dominant, impacting observables tied to the differential decay rate and thereby constraining the theoretically reliable analysis window.

A detailed comparison of the branching ratios, $A_{FB}$, $f_{L}$, and the angular coefficients $\langle I_i \rangle$ allows the identification of $q^{2}$ intervals in which theoretical uncertainties are minimized and SM predictions remain under good control. These clean windows differ among the observables and are crucial for isolating regions suitable for phenomenological studies and potential NP searches. The behavior observed in both the muon and tauon channels further emphasizes the necessity of a consistent treatment of WA contributions for precise theoretical predictions in $B_c$ semileptonic rare decays.

Overall, this study provides a complete SM baseline incorporating PB, WA, and LD contributions, and highlights several observables that remain theoretically robust within specific kinematic domains. Future progress will benefit from improved determinations of hadronic form factors and a more refined treatment of LD resonance effects, which will enhance the precision of predictions and facilitate meaningful comparisons with upcoming experimental measurements.

\bigskip

\section*{Acknowledgments}
M.A.P and Z.A would like to acknowledge the financial support provided by the Higher Education Commission (HEC) of Pakistan
through Grant no. NRPU/20-15142.

\appendix
\section{WA Form Factors}
\label{appA}
\begin{figure}[h]
    \centering
    \includegraphics[width=0.45\linewidth]{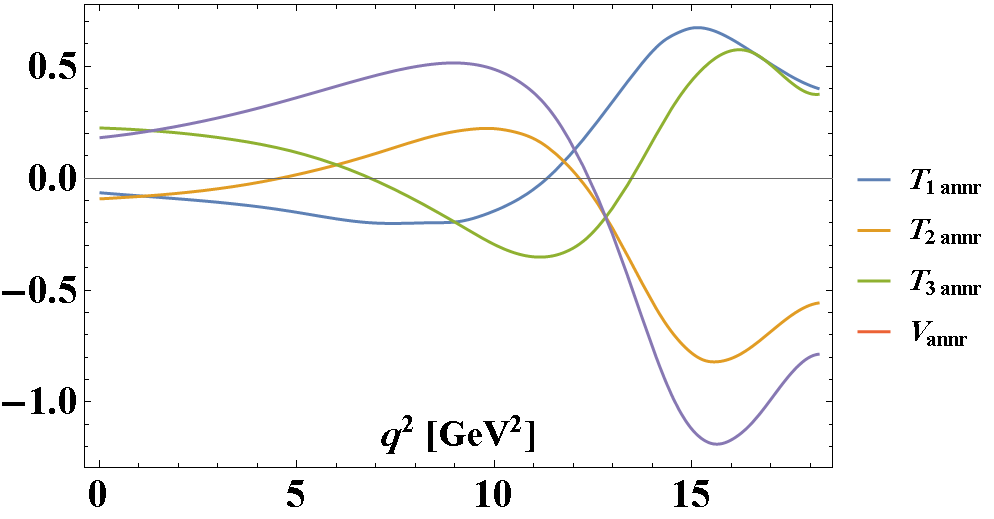}
    \includegraphics[width=0.45\linewidth]{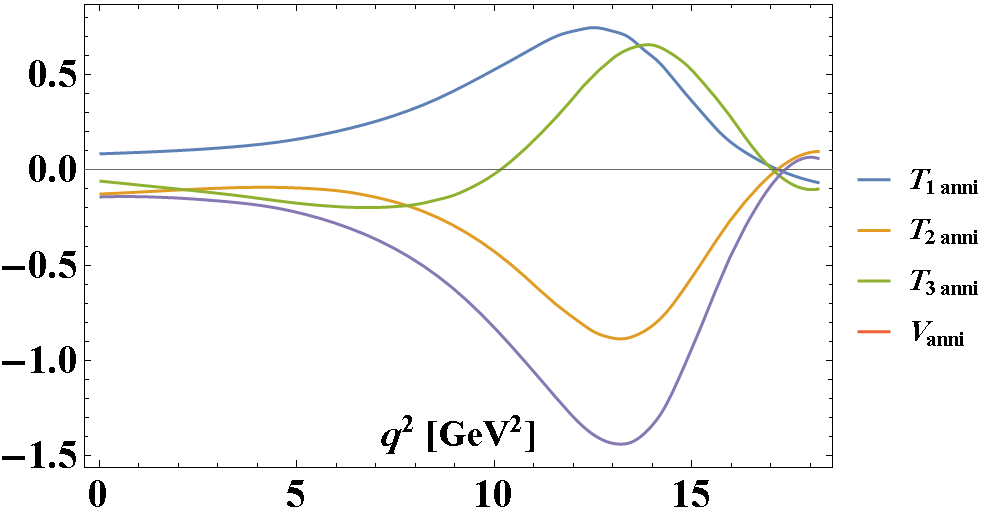}
    \caption{The real (left panel) and the imaginary (right panel) parts of the WA form factors. The uncertainties are taken to be $\pm~10\%$ which cover the uncertainty of digitization and extrapolation.}
    \label{FFsplines}
\end{figure}

The original work \cite{Ju:2013oba} provides the WA form factors $T_{1ann},~T_{2ann},~T_{3ann}$ and $V_{ann}$ in graphical form. Therefore, we digitized the published plots to reconstruct the numerical values of the form factors using the method prescribed in \cite{article}. The curves are digitized using point-selection along the graphical contours and the extracted discrete points are then interpolated using a spline interpolation.
This approach reproduces the published shapes with high accuracy and allows us to extract the form factors at arbitrary $q^2$ values as required for the numerical analysis. The plot range in the original work is $q_{max}=17$, while we have extrapolated the curves up to $q_{max}\approx18.18$. This extrapolation brings some uncertainties which are harmless, as our analysis remains in the bin $[15-17]~\text{GeV}^2$, and above this range all the curves always overlap. The reproduced results are presented in FIG. \ref{FFsplines}. 

\section{Wilson Coefficients Expressions}
\label{appB}
The explicit form of WCs \(C_{7}^{\rm eff}(q^{2})\) and \(C_{9}^{\rm eff}(q^{2})\) is given in \cite{Du:2015tda}. For the sake of completeness, we give the expressions of these WCs used in our study:

\begin{align*}
C_{7}^{\rm eff}(q^{2}) &= C_{7}-\frac{1}{3}\left(C_{3}+\frac{4}{3}C_{4}+20C_{5}+\frac{80}{3}C_{6}\right) -\frac{\alpha_{s}}{4\pi}\left[(C_{1}-6C_{2})F^{(7)}_{1,c}(q^{2})+C_{8}F^{(7)}_{8}(q^{2})\right], \\
C_{9}^{\rm eff}(q^{2}) &= C_{9}+\frac{4}{3}\left(C_{3}+\frac{16}{3}C_{5}+\frac{16}{9}C_{6}\right) -h(0,q^{2})\left(\frac{1}{2}C_{3}+\frac{2}{3}C_{4}+8C_{5}+\frac{32}{3}C_{6}\right) \\
&\quad -h(m^{\rm pole}_{b},q^{2})\left(\frac{7}{2}C_{3}+\frac{2}{3}C_{4}+38C_{5}+\frac{32}{3}C_{6}\right) +h(m^{\rm pole}_{c},q^{2})\left(\frac{4}{3}C_{1}+C_{2}+6C_{3}+60C_{5}\right) \\
&\quad -\frac{\alpha_{s}}{4\pi}\left[C_{1}F^{(9)}_{1,c}(q^{2})+C_{2}F^{(9)}_{2,c}(q^{2})+C_{8}F^{(9)}_{8}(q^{2})\right]-\alpha_s \frac{\alpha_s}{4\pi}\lambda_u^{(q)} \left[ \left( C_1 ( F^{(9)}_{1,c} - F^{(9)}_{1,u} ) \right) + C_2 ( F^{(9)}_{2,c} - F^{(9)}_{2,u} ) \right]\\
&+ Y_{res}(m_b^{pole},q^2),
\end{align*}

where the functions \(h(m^{\rm pole}_{q},q^{2})\) with \(q=c,b\), and functions \(F^{(7,9)}_{8}(q^{2})\) are defined in \cite{Beneke:2001at}, while the functions \(F^{(7,9)}_{1,c}(q^{2})\), \(F^{(7,9)}_{2,c}(q^{2})\) are given in \cite{Asatryan:2001zw} for low \(q^{2}\) and in \cite{Greub:2008cy} for high \(q^{2}\). The quark masses appearing in all of these functions are defined in the pole scheme.

\begin{table}[H]
\centering
\caption{The masses, total widths and dilepton widths of the intermediate vector mesons used in $Y_{res}(m_b^{pole},q^2)$. 
These values are quoted from the PDG~\cite{ParticleDataGroup:2022pth}.}
\begin{tabular}{lccc}
\hline\hline
$V_i$ & $m_{V_i}$ (GeV) & $\Gamma_{V_i}$ (MeV) & $B(V_i \to \ell^+ \ell^-)$ \\
 &  &  & where $\ell = e, \mu$ \\
\hline
$\rho$   & 0.775 & 149   & $4.635 \times 10^{-5}$ \\
$\omega$ & 0.783 & 8.68  & $7.380 \times 10^{-5}$ \\
$\phi$   & 1.019 & 4.249 & $2.915 \times 10^{-4}$ \\
$J/\psi$ & 3.097 & 0.093 & $5.966 \times 10^{-2}$ \\
$\psi(2S)$ & 3.686 & 0.294 & $7.965 \times 10^{-3}$ \\
\hline\hline
\end{tabular}
\end{table}

\section{Kinematics}
\label{appC}
The decay $M_{\text{in}} \to M_f \ell^+ \ell^-$ can be conveniently regarded as a quasi-two-body decay with 
$M_{\text{in}} \to M_f j_{\text{eff}}$ followed by $j_{\text{eff}} \to \ell^+ \ell^-$, where effective current 
$j_{\text{eff}}$ represents the off-shell boson. The polarization vectors of $j_{\text{eff}}$ satisfy the orthonormality 
and completeness relations as discussed in section~4. With 
$M_{\text{in}}(p) \to M_f(k) \,(j_{\text{eff}}(q) \to \ell^+(p_1)\ell^-(p_2))$, we define momenta in the rest frame of 
the parent particle $M_{\text{in}}$ as
\begin{equation}
p^\mu = (m_{\text{in}}, 0, 0, 0), \qquad 
k^\mu = (E_f, 0, 0, -|\vec{k}|), \qquad 
q^\mu = (q^0, 0, 0, +|\vec{k}|),
\end{equation}
where we choose daughter particle $M_f$ to be moving along the negative $z$ direction, and
\begin{equation}
q^0 = \frac{m_{\text{in}}^2 - m_f^2 + q^2}{2 m_{\text{in}}}, \qquad 
E_f = \frac{m_{\text{in}}^2 + m_f^2 - q^2}{2 m_{\text{in}}}, \qquad 
|\vec{k}| = \frac{\sqrt{\lambda(m_{\text{in}}^2, m_f^2, q^2)}}{2 m_{\text{in}}},
\end{equation}
where $\lambda(m_{\text{in}}^2, m_f^2, q^2)$ is the Källén function
\begin{equation}
\lambda(a, b, c) = a^2 + b^2 + c^2 - 2(ab + ac + bc).
\end{equation}

In the dilepton rest frame, considering $j_{\text{eff}}$ decaying in the $x - z$ plane, and 
$\ell^+(p_1)$ lepton making angle $\theta_\ell$ with the $z$--axis (see FIG.~\ref{cascadeDecay}),

\begin{equation}
p_1^\mu = (E_\ell, |\vec{p}_\ell| \sin\theta_\ell, 0, |\vec{p}_\ell| \cos\theta_\ell),
\end{equation}
\begin{equation}
p_2^\mu = (E_\ell, -|\vec{p}_\ell| \sin\theta_\ell, 0, -|\vec{p}_\ell| \cos\theta_\ell),
\end{equation}

with
\begin{equation}
E_\ell = \frac{\sqrt{q^2}}{2}, \qquad 
|\vec{p}_\ell| = \frac{\sqrt{q^2}}{2} \, \beta_\ell, \qquad 
\beta_\ell = \sqrt{1 - \frac{4 m_\ell^2}{q^2}}.
\end{equation}

In the $M_{\text{in}}$ rest frame, the polarization four-vectors of the effective current 
($j_{\text{eff}}$), that decays to dilepton pair are
\begin{equation}
\varepsilon^\mu(t) = \frac{1}{\sqrt{q^2}} (q^0, 0, 0, |\vec{k}|), \qquad
\varepsilon^\mu(\pm) = \frac{1}{\sqrt{2}} (0, \mp 1, -i, 0), \qquad
\varepsilon^\mu(0) = \frac{1}{\sqrt{q^2}} (|\vec{k}|, 0, 0, q^0),
\end{equation}
and in the dilepton pair rest frame the transverse polarizations of $j_{\text{eff}}$ remain same,
while the time-like and longitudinal polarizations read
\begin{equation}
\varepsilon^\mu(t) = (1, 0, 0, 0), \qquad 
\varepsilon^\mu(0) = (0, 0, 0, 1).
\end{equation}

Similarly, when the final state is vector or axial-vector particle, the polarization four-vectors 
of $V(A)$ state moving along the negative $z$ direction, in the $M_{\text{in}}$ rest frame are
\begin{equation}
\bar{\epsilon}^\mu(\pm) = \frac{1}{\sqrt{2}} (0, \pm 1, -i, 0), \qquad
\bar{\epsilon}^\mu(0) = \frac{1}{m_f} (|\vec{k}|, 0, 0, E_f).
\end{equation}

Transverse polarizations of $V(A)$ in its own rest frame remain same, whereas the longitudinal 
polarization reads
\begin{equation}
\bar{\epsilon}^\mu(0) = (0, 0, 0, -1).
\end{equation}

\providecommand{\href}[2]{#2}\begingroup\raggedright\endgroup

\end{document}